\documentclass[11pt, a4paper]{elsarticle}

\pdfoutput=1 

\usepackage{lineno,hyperref}
\usepackage[intlimits]{amsmath}
\usepackage{amssymb}

\modulolinenumbers[5]

\journal{Astroparticle Physics}

\def\simle{\lower 2pt \hbox {$\buildrel < \over {\scriptstyle \sim }$}}
\def\simge{\lower 2pt \hbox {$\buildrel > \over {\scriptstyle \sim }$}}

\def\ee{E_{\rm e}}
\def\ep{E_{\rm p}}

\def\dnp{\frac{{\mathrm d}N_{\rm p}}{\mathrm{d}p_{\rm p}}}
\def\dne{\frac{{\mathrm d}N_{\rm e}}{\mathrm{d}p_{\rm e}}}
\def\kep{K_{\rm ep}}
\def\ktildeep{\widetilde{K}_{\rm ep}}
\def\figheight{.35\textheight}

\begin{document}

\begin{frontmatter}

\title{On the non-thermal electron-to-proton ratio at cosmic
  ray acceleration sites}

\author[TPIV]{Lukas Merten\corref{corrauthor}}
\ead{lukas.merten@rub.de}
\cortext[corrauthor]{Corresponding author}

\author[TPIV]{Julia Becker Tjus}
\ead{julia.tjus@rub.de}

\author[TPIV]{Bj\"orn Eichmann}
\ead{eiche@tp4.rub.de}

\author[AIRUB]{Ralf-J\"urgen Dettmar}
\ead{dettmar@astro.rub.de}

\address[TPIV]{Ruhr Astroparticle and 
Plasma Physics Center (RAPP Center), Ruhr-Universit\"at 
Bochum\\Institut f\"ur Theoretische Physik IV / Plasma-Astroteilchenphysik\\ Universit\"atsstrasse 
150, 44801 Bochum, Germany}

\address[AIRUB]{Ruhr Astroparticle and 
Plasma Physics Center (RAPP Center), Ruhr-Universit\"at 
Bochum\\Astronomisches Institut\\ Universit\"atsstrasse 150, 
44801 Bochum, Germany}

\begin{abstract}
  The luminosity ratio of electrons to protons as it is produced in stochastic acceleration 
processes in cosmic ray sources is an important quantity relevant for several aspects of the 
modeling of the sources themselves. It is usually assumed to be around $1:100$ in the case of 
Galactic sources, while a value of $1:10$ is typically assumed when describing extragalactic 
sources. It is supported by observations that the {\it average} ratios should be  close to these 
values. At this point, however, there is no possibility to investigate how each individual source 
behaves. When looking at the physics aspects, a $1:100$ ratio is well supported in 
theory when making the following assumptions: (1) the total number of electrons and protons that is 
accelerated are the same; (2) the spectral index of both populations after acceleration is 
$\alpha_e=\alpha_p\approx 2.2$. In this paper, we reinvestigate these assumptions. In particular, 
assumption (2) is not supported by observational data of the sources and PIC simulation yield 
different spectral indices as well.
We present the detailed calculation of the electron-to-proton ratio, dropping the assumption of 
equal spectral indices. We distinguish between the ratio of luminosities and the ratio of the 
differential spectral behavior, which becomes necessary for cases where the spectral indices of the 
two particle populations are not the same. We discuss the possible 
range of values when allowing for 
different spectral indices concerning the spectral behavior of electrons and protons. 
Additionally, it is shown that the minimum energy of the accelerated population can have a large 
influence on the results. We find, in the case of the classical minimum energy of $T_{0, {\rm 
e}}=T_{0, {\rm p}}=10\,$ keV, that 
when allowing for a difference in the spectral indices of up to $0.1$ with absolute spectral indices 
varying between $2.0<\alpha<2.3$, the luminosity ratio varies 
between $0.008<\kep<0.12$. The differential particle number ratio is in the range $0.008 <\ktildeep 
<0.25$ and depends on the energy. 
\end{abstract}


\end{frontmatter}


\newpage
\section{Introduction\label{intro:sec}}

For the last couple of years, the search for the origin of cosmic rays, from GeV-energies up to
super-EeV events has started to make rapid progress. First Supernova
remnants have been identified as hadronic sources \cite{fermi_w44,fermi_ic443,tobias_sugar2015} and 
first evidence for an astrophysical high-energy neutrino
signal has been announced recently \cite{icecube_evidence2013}. The results from gamma-rays and 
neutrinos are
extremely important first steps in order to have a full identification
of the sources for the entire diffuse cosmic ray flux.
The theoretical interpretation of the signatures crucially rely on the precise modeling of the 
sources. This concerns both the prediction of the signal of gamma-ray and neutrino sources, as well 
as the interpretation of the spectral energy distribution of sources with dominant non-thermal 
signatures.

One central ingredient for these calculations is the ratio between cosmic ray electrons and 
protons. 
The ratio is typically assumed to be fixed to a one hundred times higher proton than 
electron luminosity for galactic sources (see e.g.\ 
\cite{beck_krause2005}). When discussing the sources of ultra high energy cosmic 
rays, which can accelerate particles up to $10^{21}$~eV, see e.g.\ 
\cite{waxman_bahcall1997,becker2014}, it is assumed that the electron luminosity is somewhat 
higher  than for galactic sources, i.e.\ electron to proton luminosity is 1:10. These values, that are on average supported 
by astrophysical 
observations, can also be derived following a theoretical argument also described in e.g.\ 
\cite{schlickeiser2002}:

Stochastic acceleration predicts a power-law behavior in
momentum,
\begin{equation}
\frac{dN_i}{dp_i}\propto p_{i}^{-\alpha_{i}}
\end{equation}
with $i=e,\,p$ (electrons or protons). It is now assumed that the same
total number of electrons and protons are accelerated, $N_p=N_e$ with
$N_i=\int dN_i/dT_i\,dT_i$. Here, $T_i=\sqrt{m_{i}^{2}\cdot c^4+p_{i}^{2}\cdot
  c^2}-m_{i}\cdot c^2$ is the kinetic energy of the particles with a
minimum kinetic energy of $T_0=10$~keV. Assuming further that the two
populations have the same spectral index
$\alpha_e=\alpha_p=:\alpha$, the expected ratio scales with the masses of the
two species:
\begin{equation}
\frac{dN_{e}/dp_{e}}{dN_{p}/dp_{p}}=\left(\frac{m_e+\frac{T_0}{2\cdot
    c^2}}{m_p+\frac{T_0}{2\cdot 
c^2}}\right)^{(\alpha-1)/2}\approx\left(\frac{m_e}{m_p}\right)^{(\alpha-1)/2} \label{eq:ApproxdN}
\end{equation}
The last step requires $T_0\ll m_e\cdot c^2,\,m_p\cdot c^2$. This makes the approximation in Eq.\ \ref{eq:ApproxdN} independent of the minimum energy. In contrast to that the exact solution depends strongly on the minimum energy and a possible difference in the minimum energies for protons and electrons as we will show later. With a
spectral index of $\alpha\approx 2.2$, which is an approximate value
to be expected from diffusive shock acceleration (see e.g.\ \cite{schlickeiser2002} for a summary), 
one
finds a ratio of the order of 
\begin{equation}
\ktildeep:=\frac{dN_{e}/dp_{e}}{dN_{p}/dp_{p}}\approx \frac{1}{100} \quad .
\label{diffratio:equ}
\end{equation}
Note that for equal spectral indices, $\alpha_e=\alpha_p=\alpha$, the
value remains the same for each value of $p_e=p_p$ and it is independent of momentum and thus of 
energy. Once
the indices differ from each other even slightly, this is not the case.
As a physical measure, two approaches can be pursued: either, in order to be independent of the 
energy scale, 
the total luminosities of electrons and protons can be compared:
\begin{equation}
\kep:=L_e/L_p \label{eq:DefKep}
\end{equation}
with
\begin{equation}
L_{\rm
  i}=\int_{T_{0,i}}^{T_{\max,i}}dT_{i}\,\frac{dN_i}{dT_i}\cdot T_i\,.
\end{equation}
Alternatively, the differential number ratio as defined in Eq.\ (\ref{diffratio:equ}) can be used. 
Here, it needs to be reviewed carefully for each case at what energies
the two particle populations are observed.

Considering the observation of electrons and hadrons that presumably
originate in our own Galaxy, i.e.\ cosmic rays from below the knee and
directly observed electrons, the ratio for the total luminosities of
the two particle populations comes very close to $1:100$. For
extragalactic sources, however, the comparison between the central source
candidates (Active Galactic Nuclei (AGN) and Gamma-ray bursts (GRBs))
rather suggest a ratio of $1:10$ \cite{waxman_bahcall1997}. These back-of-the
envelope calculations have their pitfalls as well, of course, as they rely on the comparison of the 
spectra after transport, including all loss processes. When an integral over source regions in which losses are differentially important the effective spectral index for the electrons can be steeper by a factor of 1 compared to protons from the same region (\citep{Kardashev1962}). In particular when concerning the electron 
spectra, that means that a fraction of the total number of particles is actually lost as they enter 
a non-relativistic regime and the numbers are not easily comparable with the calculation presented 
above. Even without  these difficulties, the electron-to-proton fraction that is observed strongly 
depends on the choice of the lower integration limit. As the propagated electron and proton spectra 
naturally have very different spectral behavior, the luminosities are compared rather than the 
differential values. This adds a further uncertainty in the calculation. These considerations 
emphasize that also
from the observational point of view, the number ratios of $1:100$ or $1:10$ for extragalactic 
sources needs to be treated with
care. One prominent example is the choice of lower integration limit for the luminosity of ultra 
high energy cosmic rays. In order to obtain the ratio $1:10$, it is assumed that the lower 
threshold 
is at the ankle, i.e.\ at $E_{\min}=10^{18.5}$~eV. It can, however, be possible that there is a 
significant part of the UHECR source flux even below the ankle, as also discussed by 
\cite{ahlers2011}, which would enhance the typically assumed ratio. If we assume $E_{\min}=10^{17.3}$~eV as the beginning of the extragalactic part of the spectrum as the KASCADE Grande data \cite{Apel2013, Apel2014} suggests the ratio would decrease to $1:25$. This is discussed in more detail in section \ref{ssec:ExtragalacticSources}.

From the theoretical point of view, first results from PIC simulations show that the acceleration 
of 
protons and electrons in shock fronts yields differences in the spectral behavior 
\cite{sironi_spitkovsky2011}. If the acceleration itself does not only depend on the charge but 
also on the Larmor radius of the particles such differences are expected. Radio observations of 
SNRs can also be used to show that the electron and proton spectra at 
the source differ significantly from each other \cite{snr_matthias}. However, this concerns the 
loss-dominated electrons and it is not trivial to compare these values to the spectra 
immediately after acceleration, which is needed as an input for our calculations (see also 
\cite{eichmann_beckertjus2016}).

Given the arguments from above, we revisit the theoretical calculation of the
electron-to-proton ratio in more general terms as was done before in
order to examine the possible range for individual source classes. The
assumptions we use are the following:
\begin{enumerate}
\item We assume a power-law behavior in momentum for both species, 
\begin{equation}
\frac{dN_i}{dp_{i}}=A_i\cdot p^{-\alpha_i} \label{eq:powerLaw}\,.
\end{equation}
This type of spectral behavior is expected from diffusive shock acceleration processes and is in 
agreement with the observed spectrum of leptonic and hadronic cosmic rays, see e.g.\ 
\cite{gaisser1990} for a review.
\item We
drop the assumption of equal spectral indices
($\alpha_e\neq \alpha_p$), as the acceleration process itself may depend 
on the particle masses as suggested in PIC simulations (see e.g.\ \cite{Sironi2011}).
This implies that the ratio of electrons to
protons becomes energy
dependent when considering it in $dN_i/dp_i$: $\ktildeep=\ktildeep(\ep,\,\ee)$. We therefore also 
calculate
the ratio of total luminosities $\kep$ as a true observational measure for
the electron-to-proton ratio given in Eq.\ \ref{eq:DefKep}.
\item We assume that the total number of particles accelerated in a source is the same for
  electrons and protons:
\begin{equation}
N_{p}=N_{e}\,.
\end{equation}
This assumption is supported by the following argument: The overall particle number, accelerated and non-accelerated, is the same for protons and electrons due to charge balance $N_{\mathrm tot, p}=N_{\mathrm tot, e}$. If we assume the particles to be in a thermal equilibrium in the absence acceleration, the number of particles $N_i$ above a certain energy threshold $T_0$ is the same for protons and electrons. A plasma in a thermal equilibrium is described by a Maxwellian distribution with equal temperature for protons and electrons, which leads to:
\begin{equation}
N_i = N_{tot, i}\cdot\int_{T_0}^\infty 2 \sqrt{E/ \pi} (k_b T)^{3/2} \exp{\left( -\frac{E}{k_b T} \right)} \mathrm{d}E \, . \label{eq:Maxwell}
\end{equation}
Therefore, the number of particles above threshold energy $T_0$ is independent of the particle mass.
\item We use a general lower kinematic energy threshold which is not necessarily the same for
  protons and electrons, $T_{0,e}\neq T_{0,p}$. The value of this lower kinematic threshold depends on the theory that is used to describe the acceleration process and does not necessarily describe the boundary between the thermal and non-thermal populations. The derivation of a consistent value for the minimum energy is beyond the scope of this paper but some ideas are discussed in section \ref{results:sec}. In the case of unmodified Maxwellian distributions this would introduces a charge imbalance for the accelerated particles (see Eq. \ref{eq:Maxwell}). Different acceleration mechanisms for electrons and protons due to other plasma wave scattering processes might also lead to different numbers of accelerated particles. In a generalized approach such a charge imbalance could be treated consistently via the introduction of a new parameter $\eta$:
\begin{equation}
  \eta=\frac{N_e}{N_p}= \frac{N_{tot, e}-N_{low, e}}{N_{tot, p}-N_{low, p}}\\
  = (1-\int_{0}^{T_{0, e}} P_e(E) \mathrm{d}E) / (1-\int_{0}^{T_{0, p}} P_p(E) \mathrm{d}E) \label{eq:eta} \, ,
\end{equation}
where $P_i$ describes the energy density distribution below the threshold energy for effective acceleration $T_{0,i}$.

\end{enumerate}

In previous work, \cite{persic_rephaeli2014} apply an approach similar to ours, dropping the 
assumption of equal indices. There are several differences in this work as compared to 
\cite{persic_rephaeli2014}:
(1) While \cite{persic_rephaeli2014} focus on starburst galaxies in their values for the 
calculation, we systematically investigate the electron-to-proton ratio for sources up to the 
cosmic 
ray knee ($E_{\max} \sim 10^{15}$~eV) and those that contribute up to the end of the cosmic ray 
spectrum at $10^{21}$~eV. (2) We discuss both differential ratios for the energy-dependent spectra 
and the ratio of the luminosities. (3) We present an analytical solution to the calculation which 
can be applied to calculations for cosmic ray source candidates in the future. This should be 
possible for most sources, as we do not use any specific information for any source class in order 
to keep the approach as general as possible. (4) We allow for different minimum energies $T_{0, 
{\rm i}}$ which accounts for different effective acceleration thresholds for protons and electrons 
due to their different gyro-radii (see e.g.\ \cite{Bell1978_I, Bell1978_II}).

This paper is organized as follows: In Section \ref{calc:sec}, the
calculation of the electron-to-proton ratio is presented. In contrast to most of the previous 
calculations, we 
perform a full 
calculation of the total luminosities in electrons and protons as a physical measure, rather than 
the approximation of
the ratio of differential spectra at a fixed energy.
The central theoretical results in terms of
quantitative deviation from the typically used approximation of equal
indices are discussed in Section \ref{results:sec}. In Section \ref{discussion:sec}, we discuss how 
our result matches current observations concerning potential
extragalactic and Galactic cosmic ray sources. We also discuss the 
implications of this result on the interpretation of neutrino,
gamma-ray and magnetic field data within current astrophysical cosmic ray interaction models. In 
Section \ref{sum:sec}, we summarize our result and give an
outlook to future applications.

\newpage

\section{Acceleration theory: calculation of the electron-to-proton ratio \label{calc:sec}}
In most literature the ratio of differential particle numbers 
$\frac{\mathrm{d}N_e}{\mathrm{d}p_e}\big/\frac{\mathrm{d}N_p}{\mathrm{d}p_p}$ and the ratio of the 
total luminosities $K_{ep}$ is used equivalently. In addition, a ratio of $K_{ep}=1/100$ is usually 
applied, assuming spectral indices of $\alpha_e=\alpha_p=2.2$. These assumptions are not reflected 
in reality. In \cite{snr_matthias}, for instance, it is shown that the assumption of equal spectral 
indices does not need to hold. Here, we calculate the ratio of luminosities, leading to the 
prediction of the electron-to-proton ratio for different combinations of indices.

In this section, we separate the calculation into two parts: derivation of an effective normalization ratio 
$A_{ep}=A_{\rm e}/(A_{\rm p} \eta)$ with $A_i$ as it is defined in Eq.\ \ref{eq:powerLaw} and the remaining part of 
the integral to determine the ratio of luminosities $\kep$. We further provide estimates of the 
ratio of the different particle spectra as a function of the primary energies, 
$\ktildeep(\ep,\,\ee)$.

\subsection{Normalization ratio \label{norm:subsec}}
As a first step of the calculations the ratio of normalizations constants $A_{\rm e}$ and $A_{\rm 
p}$ corrected by the factor $\eta$ is derived. We follow the calculations of \cite{schlickeiser2002}, without applying any 
approximations and allowing for different spectral indices for electrons and protons, i.e.\ 
$\alpha_e\neq \alpha_p$ but in addition we account for a possible difference in the number of accelerated particles (see Eq. \ref{eq:eta}). Using the assumption that the total particle number for protons and 
electrons is the same, one obtains:
\begin{align}
\eta\cdot N_{\rm p} &= N_{\rm e} \\
\eta\cdot\frac{A_{\rm p}}{\alpha_{\rm p}-1}(T_{0, {\rm p}}^2+2T_{0, {\rm p}}m_{\rm p})^{-\frac{\alpha_{\rm 
p}-1}{2}} &= 
\frac{A_{\rm 
e}}{\alpha_{\rm e}-1}(T_{0, {\rm e}}^2+2T_{0, {\rm e}}m_{\rm e})^{-\frac{\alpha_{\rm e}-1}{2}}\, .
\end{align}
This leads directly to the following relation which cannot be simplified any further:
\begin{equation}
A_{ep}=\frac{A_{\rm e}}{A_{\rm p}}\cdot\eta^{-1}=\frac{{(\alpha_{\rm e}-1)}(T_{0, {\rm p}}^2+2T_{0, {\rm 
p}}m_{\rm 
p})^{-\frac{\alpha_{\rm p}-1}{2}}}{{(\alpha_{\rm p}-1)}(T_{0, {\rm e}}^2+2T_{0, {\rm e}}m_{\rm 
e})^{-\frac{\alpha_{\rm 
e}-1}{2}}}\cdot\eta^{-1}\, . \label{eq:norm}
\end{equation}

\subsection{Integration\label{integral:subsec}}
The integration in (\ref{eq:Ltot2}) is performed over the kinetic energy $T$ so we start with 
rewriting the differential particle numbers.
\begin{equation}
\frac{{\rm d}N_i}{{\rm d}T_i}=\frac{\partial{N_i}}{\partial{p_i}} \cdot \frac{{\rm d}p_i}{{\rm 
d}T_i}\\
=A_{{\rm i}}\cdot(T_i+m_i)\cdot(T_i^2+2T_im_i)^{-\frac{\alpha_i+1}{2}} \label{eq:dN}
\end{equation}
Substituting relation (\ref{eq:dN}) into the formula for the total luminosity $L_{\rm tot}$ 
(\ref{eq:Ltot2}) we get:
\begin{equation}
L_{\rm
  tot,i}=\int_{T_{0,i}}^{T_{\max,i}}dT_{i}\,A_{\rm i}\cdot(T_{\rm i}+m_{\rm i})\cdot(T_{\rm 
i}^2+2T_{\rm i}m_{\rm i})^{-\frac{\alpha_{\rm i}+1}{2}}\cdot T_i := A_{\rm i} \cdot L^0_{\rm
  tot,i}\, .\label{eq:Ltot2}
\end{equation}
The integral $L^0_{\rm tot,i}$ can be separated into four parts, yielding: 
\begin{align}
L^0_{\rm tot, i}&=\int_0^{T_{\rm max, i}} {\rm d}T\, T^2\cdot(T^2+2mT)^{-a} - \int_0^{T_{0, {\rm 
i}}} {\rm 
d}T\, T^2\cdot(T^2+2mT)^{-a} \notag \\ &+ \int_0^{T_{\rm max, i}} {\rm d}T\, Tm\cdot(T^2+2mT)^{-a} 
- 
\int_0^{T_{0, {\rm i}}} {\rm d}T\, Tm\cdot(T^2+2mT)^{-a}
\end{align}
where $a=(\alpha+1)/(2)$.
This is possible in such 
a way that each of these integrals is analytically well defined and converges for nearly all 
physically interesting parameters. 
These integrals can be transformed into integrals with a well known solution in form of 
hypergeometric functions ${_2}F_1(a, b; c; z)$ \cite{ABR72}:
\begin{equation}
\int_0^1{\rm d}t\, t^{b-1}\cdot(1-t)^{c-b-1}\cdot(1-tz)^{-a} = 
\frac{\Gamma(b)\Gamma(c-b)}{\Gamma(a)}\cdot{_2}F_1(a, b; c; z)\, . \label{eq:Abramowitz}
\end{equation}
The analytic continuation of the hypergeometric function for $z\gg 1$ converges if the physical 
parameter fulfills $\alpha\notin Z$ \cite{WOL}. Subsequent to the fixing of $a,b,c$ and $z$ we 
obtain 
the following result:
\begin{align}
 L^0_{\rm tot, i}(m,a) = \frac{(2m)^{-a}}{(3-a)(2-a)}\cdot  &\left[(2-a)\cdot T_{\rm max, i}^{3-a} 
\cdot 
{_2}F_1\left(a, 3-a; 4-a; -\frac{T_{\rm max, i}}{2m}\right)\right. \notag \\
					  &+ (a-2)\cdot T_{0, {\rm i}}^{3-a}\cdot {_2}F_1\left(a, 
3-a; 4-a; 
-\frac{T_{0, {\rm i}}}{2m}\right)         \notag \\
					  &+ (3-a)\cdot mT_{\rm max, i}^{2-a}\cdot {_2}F_1\left(a, 
2-a; 
3-a; -\frac{T_{\rm max, i}}{2m}\right)        \notag \\
					  &+ \left.(a-3)\cdot mT_{0, {\rm i}}^{2-a}\cdot 
{_2}F_1\left(a, 2-a; 
3-a; -\frac{T_{0, {\rm i}}}{2m}\right) \right]\label{eq:L0tot}
\end{align}
Finally, the luminosity ratio of electrons and protons can be expressed as
\begin{align}
K_{ep}=A_{\rm ep}\cdot\frac{L^0_{\rm tot, i}(m_{\rm e},\frac{\alpha_{\rm e}+1}{2})}{L^0_{\rm 
tot, i}(m_{\rm p},\frac{\alpha_{\rm p}+1}{2})} \, .
\end{align}
It is worth noticing that no algebraic approximations are necessary to obtain this result. This is 
in contrast to the approximated ratio, based on the calculation of the comparison of the 
differential particle fluxes, where it is usually assumed that the minimum kinetic energy is much 
smaller than the particle masses, $T_0\ll m_e,\,m_p$, see e.g.\ \cite{schlickeiser2002}. When we 
calculate $\kep$ for the same parameters as given in \cite{persic_rephaeli2014} we derive very 
similar results.

\subsection{Differential particle number\label{number:subsec}}
For certain scenarios, it can be more useful to work with the ratio of the differential particle 
flux, which we define as
\begin{equation}
\ktildeep(p_e,\,p_p)=\frac{\dne(p_e)}{\dnp(p_p)}
\end{equation} 
and can be rewritten as:
\begin{equation}
\ktildeep(T_{\rm{e}}, T_{\rm{p}})=A_{{\rm ep}}\cdot \frac{T_{\rm e}+m_{\rm 
e}}{T_{\rm p}+m_{\rm p}}\cdot\frac{{(T_{\rm 
e}^2+2T_{\rm e}m_{\rm 
e})}^{-\frac{\alpha_{\rm e}+1}{2}}}{({T_{\rm 
p}^2+2T_{\rm p}m_{\rm p})}^{-\frac{\alpha_{\rm p}+1}{2}}} \quad .
\end{equation}

Proton and electron spectra are directly linked to their photon (or neutrino) emission spectra. 
Hadronic interactions produce neutral and charged pions and kaons, which lead to high-energy photon 
and neutrino emission, while synchrotron radiation, Inverse Compton processes and bremsstrahlung 
are 
the main channels for electromagnetic radiation from electrons (see \cite{snr_matthias} and 
references therein). It is often the case that broadband information is not available, but that 
correlation studies are done in a limited frequency band. As an example, for the 
mono-chromatic approximation of synchrotron radiation, 
a single observed frequency at radio wavelengths can be approximated as coming from one energy (or 
momentum $p_e$) in the energy spectrum and a similar approximation can be done in the case of the 
correlation between hadronic gamma-rays and protons (momentum $p_{\rm p}$). The choice of frequency 
in both cases depends on the available data and therefore, it is realistic that $p_e\neq p_p$.  
For each individual case, it is therefore important to 
define beforehand to which frequencies or frequency range the calculation relates. Concrete 
astrophysical examples 
are discussed in this paper in Section \ref{discussion:sec}.

\newpage

\section{Results \label{results:sec}}
In this section, we investigate the uncertainties of the ratios of electrons and protons 
considering the normalization ratio $A_{\rm e,p}$, the differential number ratio, $\ktildeep$, and 
the luminosity ratio, $\kep$. Consequences for 
the modeling of galactic and extragalactic cosmic ray sources are discussed in Section 
\ref{discussion:sec}. 

\subsection{Normalization}
\label{ssec:Normalization}
According to the common approach the minimum energy for the normalization ratio $A_{\rm e,p}$ is 
fixed to $T_0=10$~keV 
(e.g.\ \cite{schlickeiser2002, persic_rephaeli2014}). However, it is plausible that an equal minimum 
energy does not need to hold at all acceleration sides. If the minimum energy for an 
effective acceleration is connected to the gyro-radius it will differ significantly for protons and 
electrons. Bell showed in \cite{Bell1978_II} that the minimum energy for accelerated electrons 
should be higher than for protons because they are more likely scattered inside the shock front 
which prevents an efficient acceleration. It should be mentioned that other authors, like Morlino, show that also a spread of the minimum energy is possible, due to the injection of electrons coming from high energy ions \citep{Morlino2009}. Furthermore, Malkov and V\"olk did some calculations on the transition between the thermal and the accelerated spectrum based on diffusive particle acceleration at parallel shocks which might be a good starting point for further research on the transition region \cite{Malkov1995}.

Before we looked into the consequences of different minimum energies for protons and electrons we 
examined the influence of a varying but equal minimum energy. In \cite{Bell1978_II} the relation 
between the shock speed $v_{\rm shock}$ and the minimum energy $T_0$ is given by:
\begin{equation}
 T_0 = 4\cdot\left(\frac{1}{2}mv_{\rm shock}^2\right)\, .\label{eq:ShockSpeed}
\end{equation}

\begin{figure}
\centering{
\includegraphics[height=\figheight]{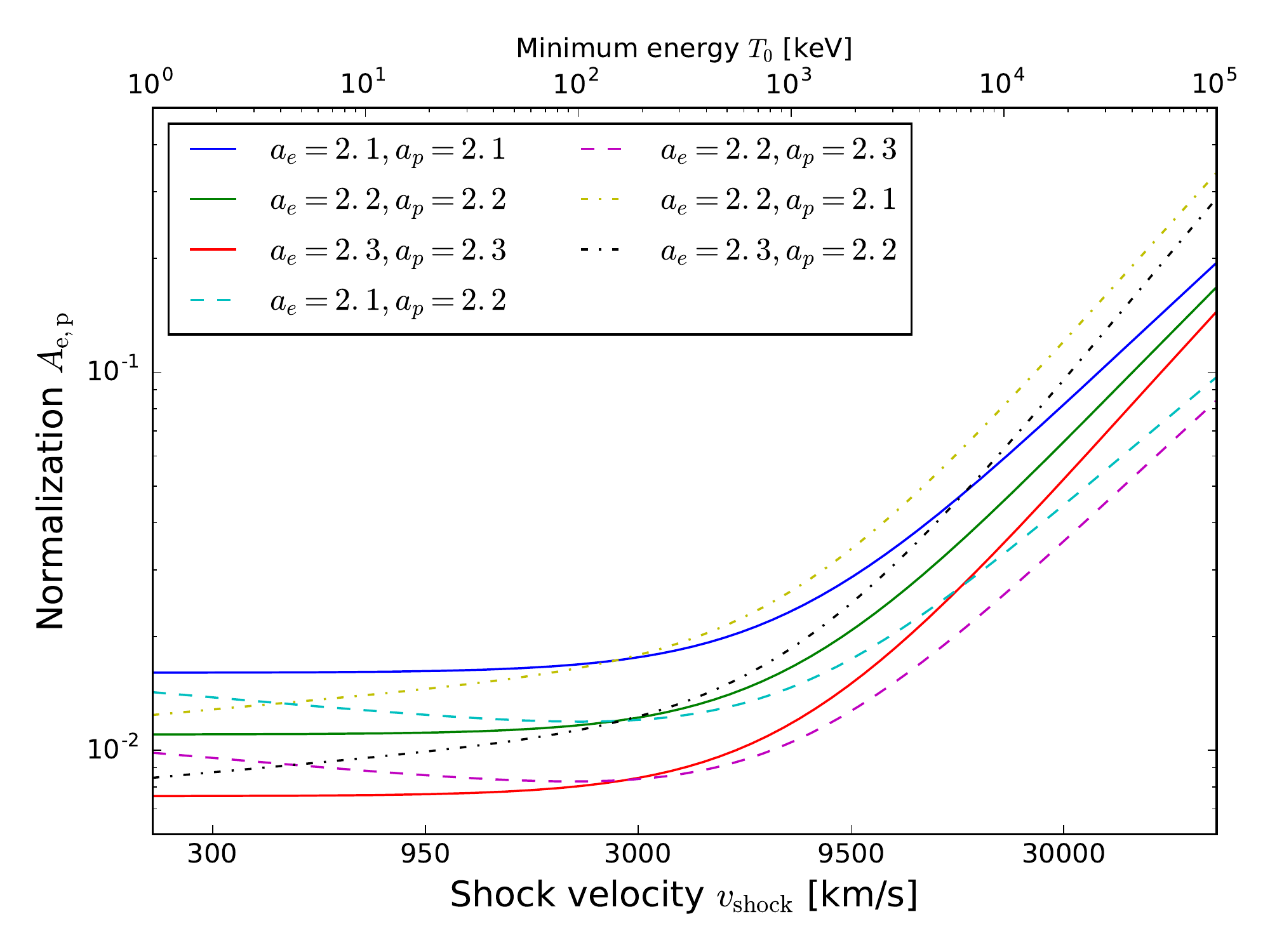}
\caption{The shock speed is calculated corresponding to \cite{Bell1978_II} and Eq. 
\ref{eq:ShockSpeed}.}\label{fig:Norm}
}
\end{figure}

In Fig.\ \ref{fig:Norm} the normalization ratio $A_{\rm ep}$ is shown for different minimum 
energies and some exemplary combinations of spectral indices. On the two axes the minimum energy 
$T_0$ and the corresponding shock speed $v_{\rm shock}$ are displayed. It is clearly visible that 
only shock velocities above a few thousand kilometers per second, $v_{\rm shock}\gtrapprox 
5000$~km/s have a significant influence on the normalization ratio. The normalization ratio is nearly 
constant for equal minimum energies up to $T_{0, {\rm i}}\leq0.1$~MeV. For higher energies the ratio 
increases with a power law behavior.

\begin{figure}
\centering{
\includegraphics[height=\figheight]{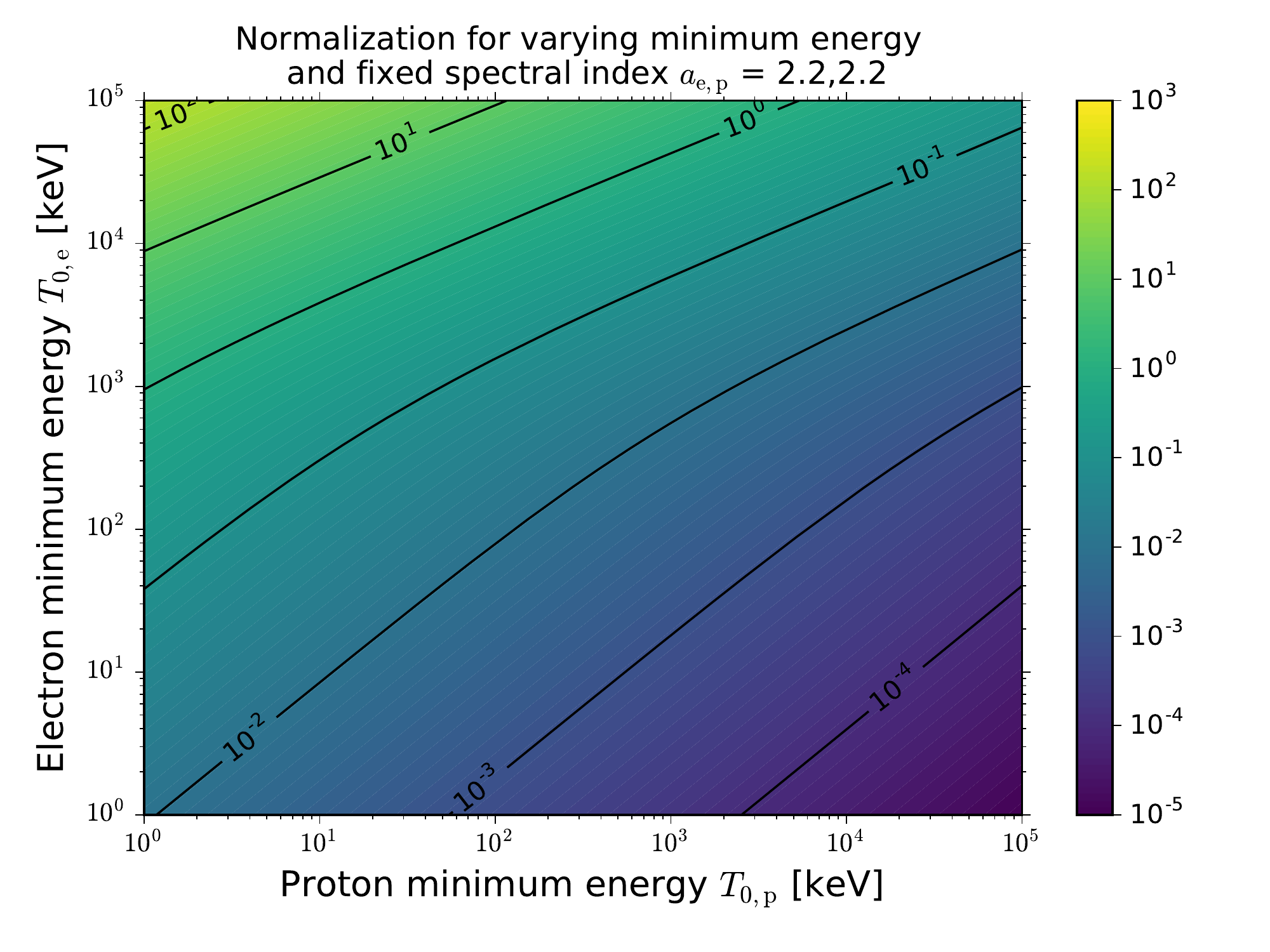}
\caption{Normalization ratio $A_{\rm e,p}$ for different minimum energies 
for protons and electrons, $T_{0,{\rm p}}\neq T_{0,{\rm e}}$.}\label{fig:Norm_Contour}
}
\end{figure}

Figure \ref{fig:Norm_Contour} shows the normalization ratio for different minimum energies. We assume, here and for all other figures and calculations, that $\eta=1$ to emphasize the influence of the different minimum energies and not the influence of different particle numbers. Here it 
is clearly visible that different acceleration mechanisms or efficiency scales have an significant 
impact on the normalization ratio. Therefore, it is mandatory to look into the details of the 
acceleration process to correctly use the formulas provided in this paper. 

For further investigations of $\kep$ and $\ktildeep$ we use the conventional minimum energy of 
$T_{0, {\rm e}}=T_{0, {\rm p}}=T_0=10$~keV, if not otherwise stated.

\subsection{Luminosity ratio $\kep$}
The luminosity ratio depends on the maximum energy used for the integration. Here, we present the 
results for galactic sources, accelerating particles up to at least the knee, i.e.\ we consider a 
maximum energy of $10^{6}$~GeV$<E_{\max}=E_{p, \max}=E_{e, \max}<10^{8}$~GeV (see e.g.\ 
\cite{biermann1993,biermann_apj2010}), as well as for extragalactic 
sources, accelerating particles to beyond the ankle, i.e.\ $E_{\max} \sim 10^{12}$~GeV. As we 
consider protons here, the acceleration might be somewhat less than $10^{11}$~GeV and we consider a 
range of $10^{8}$~GeV$<E_{\max}=E_{p, \max}=E_{e, \max}<10^{10}$~GeV to account for the fact that a 
possible present 
fraction of heavy nuclei would dominate the spectrum with their maximum energy at the highest 
energies of $10^{12}$~GeV.

\paragraph{Galactic Cosmic Rays}
\begin{figure}
\centering{
\includegraphics[height=\figheight]{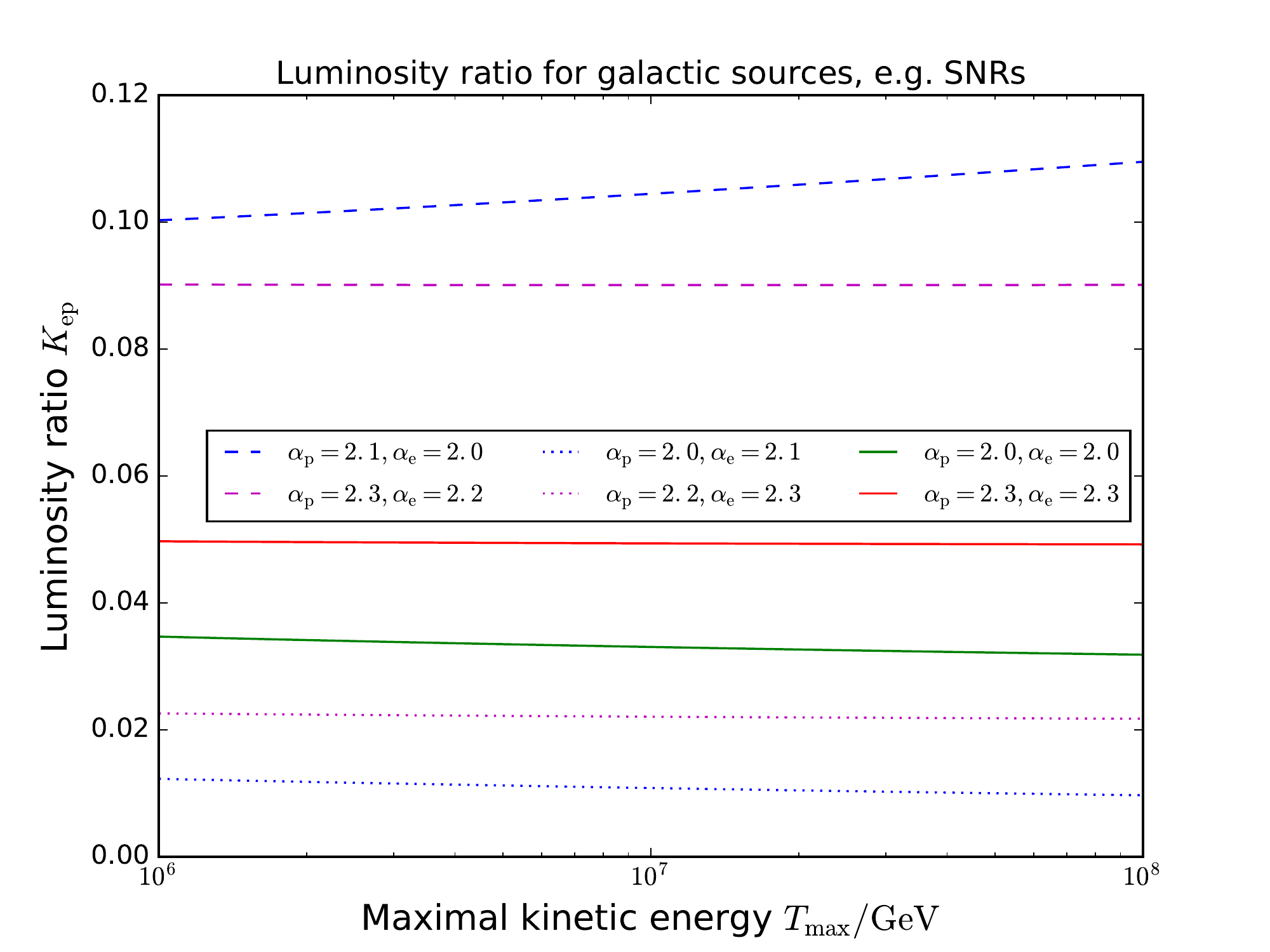}
\caption{The luminosity ratio $K_{\rm ep}$ does not dependent strongly on the maximum energy 
$T_{\rm {max}}=T_{\rm{p, max}}=T_{\rm{e, max}}$ but some differences between equal (solid lines) 
and different (dashed lines) 
indices are visible.}\label{kep_gal}
}
\end{figure}

The luminosity ratios for galactic sources are shown in Fig.\ \ref{kep_gal}, as a function of the 
maximum energy. The standard approach is to take equal spectral indices for electrons and protons, 
i.e.\ $\alpha_e=\alpha_p=\alpha$. In the figure, we show the 
variation of indices in the range $2.0<\alpha<2.3$. For equal indices, we find that the luminosity 
ratio does not vary with the maximum energy at all, but gives different values depending on the 
choice of 
the spectral index, i.e.\ between $0.04<\kep<0.07$.
When applying a slight 
difference between the spectral indices, i.e.\ $\Delta\alpha=\alpha_e-\alpha_p=-0.1$ the ratio 
becomes significantly larger and slightly energy dependent. In that case, the ratio is rather in 
the range of 
$\kep\sim 0.10-0.11$ when the kinetic energy is about $T=10^{6}-10^8$~GeV. For the opposite case, 
meaning $\Delta \alpha =+0.1$, the ratio becomes significantly smaller and the energy dependence is 
less pronounced. Here, the ratio is in the range of $\kep\sim 0.012-0.03$ when the kinetic 
energy is about $T=10^{6}-10^8$~GeV.

\begin{figure}
\centering{
\includegraphics[height=\figheight]{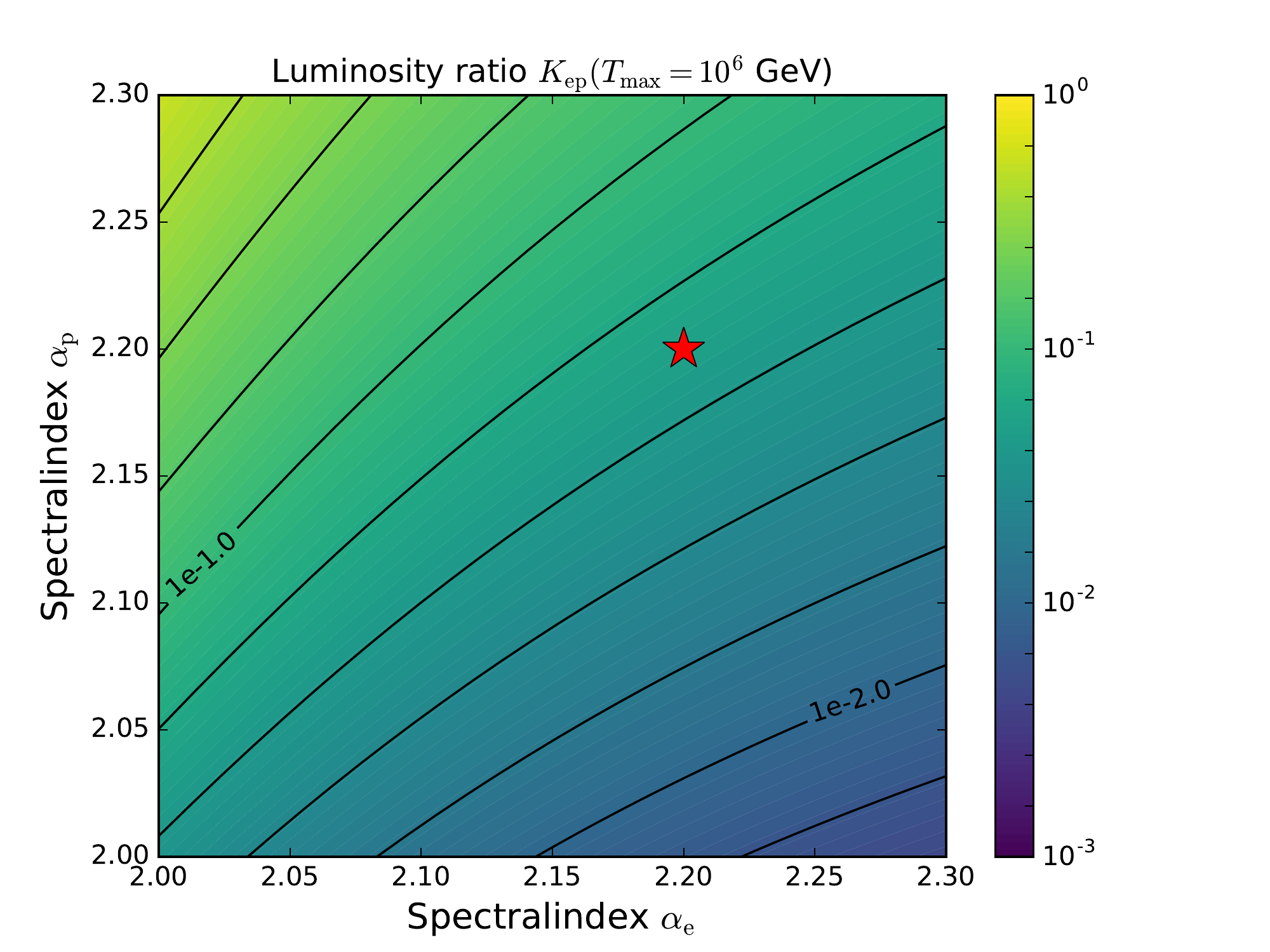}
\caption{Galactic luminosity ratio $\kep$ at a fixed maximum kinetic energy of 
$T_{\max}=10^{6}$~GeV and for arbitrary spectral index combinations in the range of  
$2.0<\alpha_{\rm e},\,\alpha_{\rm p}<2.3$. The red star indicates the 
conventional ratio with $\alpha_{\rm e, p}=2.2$.}\label{kep_gal_contour}
}
\end{figure}

Figure \ref{kep_gal_contour} shows a contour plot in more general 
terms, i.e.\ the luminosity ratio 
for arbitrary spectral index combinations for absolute values of $2.0<\alpha_e,\,\alpha_p<2.3$ for 
a maximum energy of $T=10^{6}$~GeV. Thus, the figure includes at maximum a 
relatively large difference between the spectral indices of $|\Delta\alpha|=0.3$. In these most 
extreme cases, the ratio varies between $0.004<\kep<0.6$ for $T=10^{6}$~GeV.

\paragraph{Extragalactic Cosmic Rays}

\begin{figure}
\centering{
\includegraphics[height=\figheight]{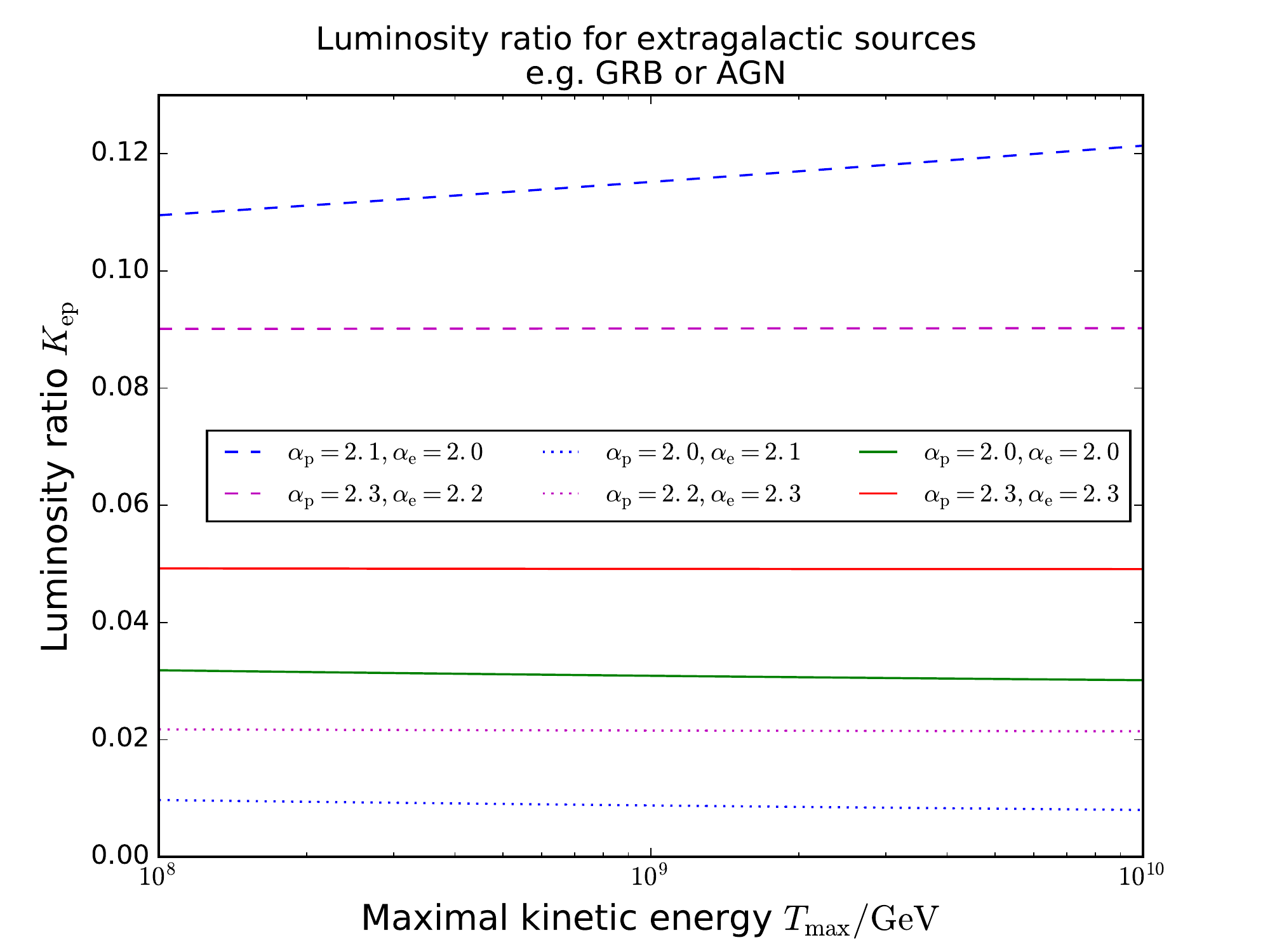}
\caption{Luminosity ratio $\kep(T_{\max})$ for different pairs of spectral indices 
$\alpha_e,\,\alpha_p$. 
Equal spectral indices (solid lines) lead to the same constant ratio as for the galactic 
energy range. A difference in the spectral indices (dashed lines) of $\delta \alpha=-0.1(+0.1)$ 
leads to a ratio about one order of magnitude higher (lower). Furthermore, some index 
combinations lead to a energy dependent behavior.}\label{kep_extragal}
}
\end{figure}
For extragalactic sources, results for the maximum energy dependent luminosity ratio are shown in 
Fig.\ \ref{kep_extragal}, in the same way as for galactic sources. Here, the luminosity ratio in 
the case of equal indices varies between $0.03<\kep<0.07$, just as for galactic sources, as there 
is no energy dependence. Again, for a difference in the spectral indices of protons and electrons of 
$\Delta\alpha=-0.1(+0.1)$, the ratio becomes larger (smaller) and energy dependent, so that it lies 
in a range $\kep\sim 0.10-0.12 (0.009-0.03)$. Due to the moderate energy dependence of $\kep$ for 
the case of $\alpha_e\neq \alpha_p$, the ratio is slightly higher for extragalactic sources as 
compared to galactic ones. 

\begin{figure}
\centering{
\includegraphics[height=\figheight]{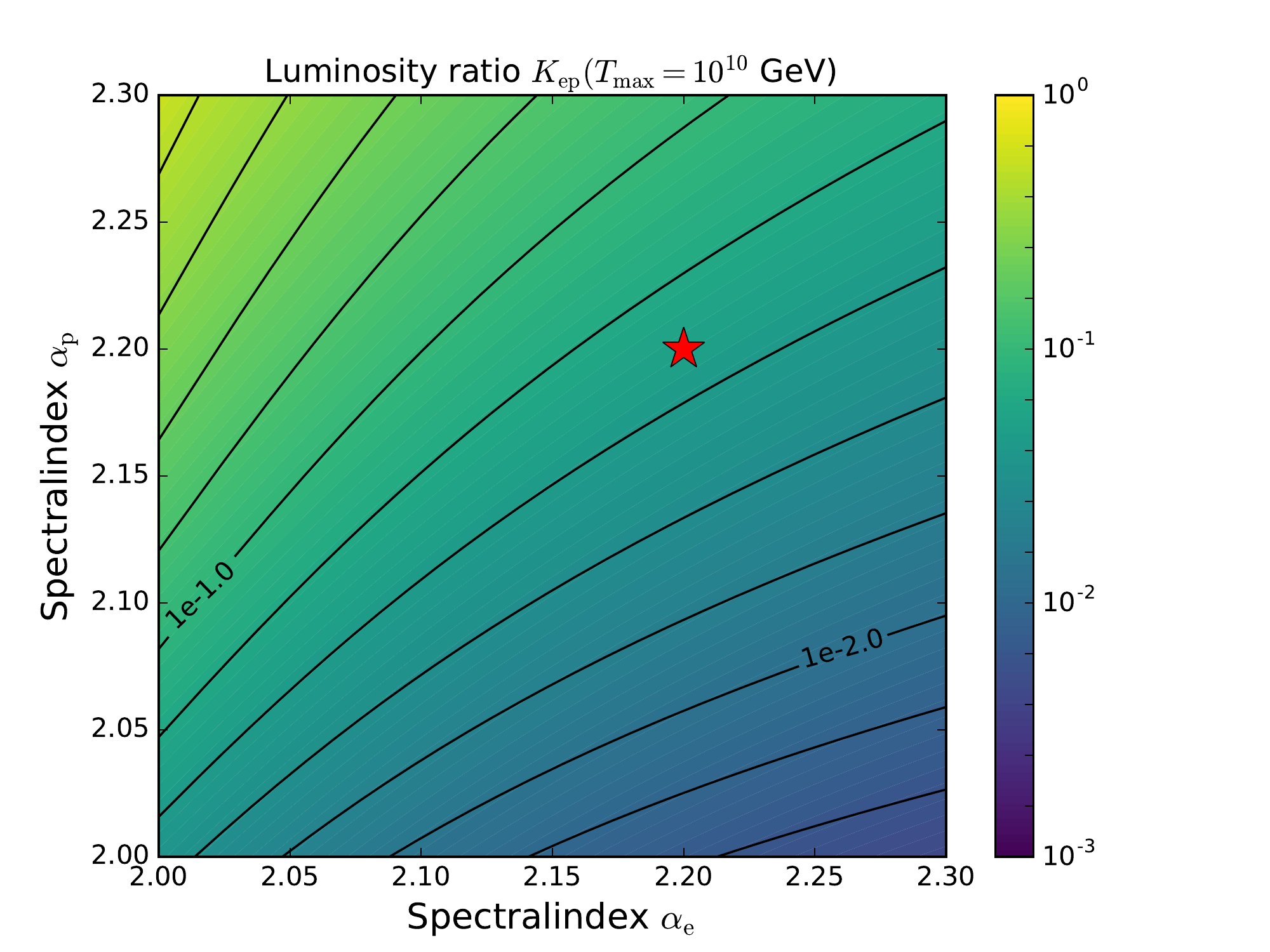}
\caption{Luminosity ratio $\kep$ at a fixed maximum kinetic energy 
of $T_{\max}=10^{10}$~GeV and for arbitrary spectral index combinations in the range of  
$2.0<\alpha_e,\,\alpha_p<2.3$. The red star indicates the 
conventional ratio with $\alpha_{\rm e, p}=2.2$.}\label{kep_extragal_contour}
}
\end{figure}
The contour plot for a fixed maximum energy of $10^{10}$~GeV and spectral indices in the range 
$2.0<\alpha_e,\,\alpha_p<2.3$ is shown in Fig.\ \ref{kep_extragal_contour}. The standard value of 
$(\alpha_e,\,\alpha_p)=(2.2,\,2.2)$ is indicated as a star. For the most extreme deviations 
of $|\Delta\alpha|=0.3$, the results range from $0.0026<\kep<0.8$. A 
difference between the indices of $0.3$ is rather large, so it is expected that the actual 
variation 
is significantly smaller, i.e.\ rather as we discuss it in Fig.\ \ref{kep_extragal}.

Here, it becomes obvious that the energy dependence of the luminosity ratio turns out to be rather 
weak so that the ranges for galactic and extragalactic sources actually are comparable.

\paragraph{Different minimum energies}

In this paragraph we show exemplarily how the luminosity ratio $\kep$ is influenced by different 
minimum energies for protons and electrons, $T_{0,{\rm p}}\neq T_{0,{\rm e}}$. Since the exact 
principle of the acceleration mechanism at the sources of cosmic rays is not yet completely 
understood and a variety of different theories exist a difference in the minimum energies is in 
principle possible.

\begin{figure}
\centering{
\includegraphics[height=\figheight]{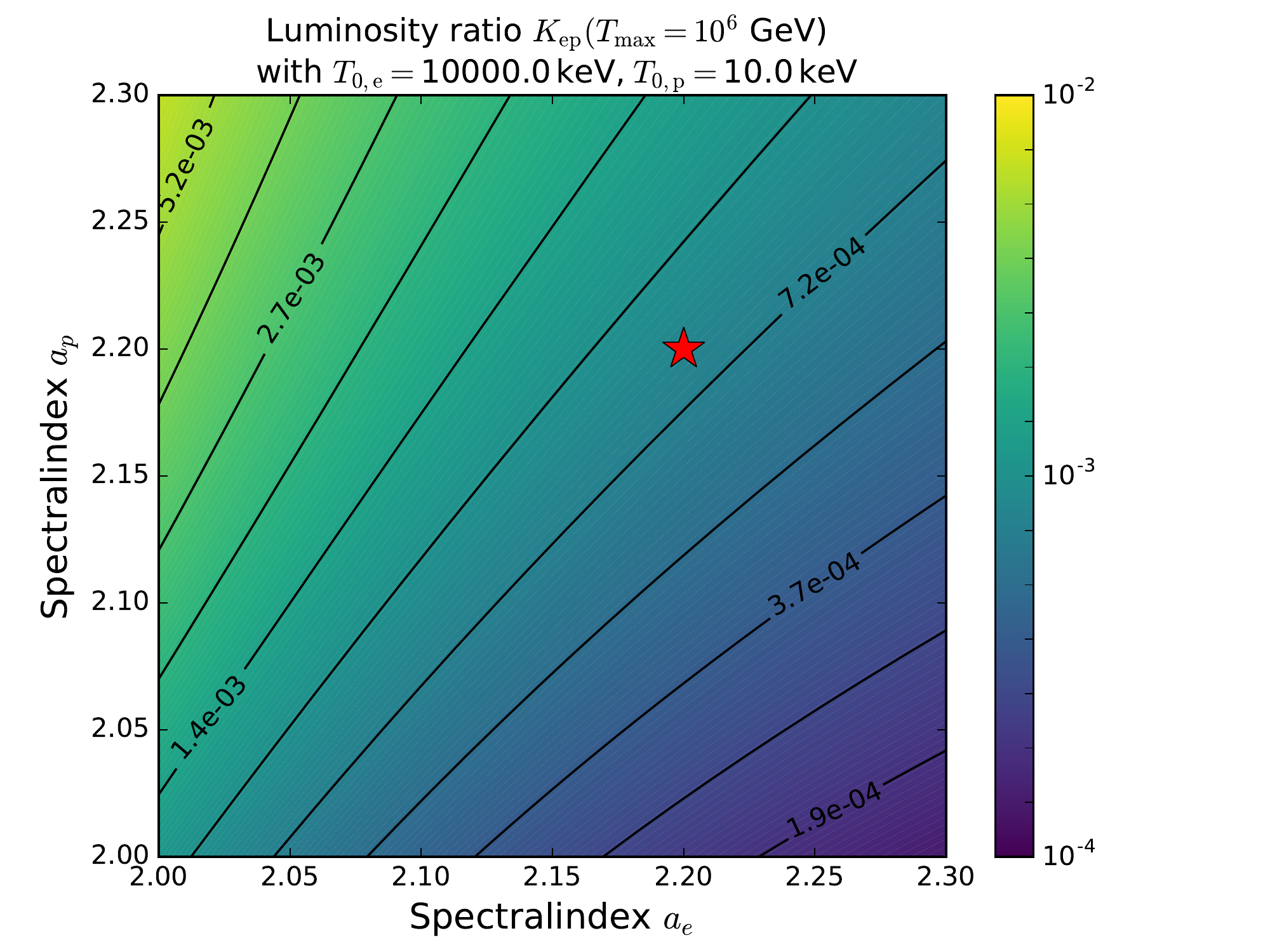}
\caption{The difference in the minimum energies shifts the whole co-domain of 
$\kep$ to lower values compared to Fig.\ \ref{kep_gal_contour}. Furthermore, 
the influence of $\alpha_{\rm e}$ is less strongly.}\label{fig:GalConDiff}
}
\end{figure}

For this example we decided to use the approach given in \cite{Bell1978_I, Bell1978_II}. Here, the 
main assumption is that particles will only be efficiently accelerated if they do scatter inside 
the 
shock front. Meaning the gyro-radius of the particles has to be larger than the thickness of the 
shock front. Due to the mass difference this leads to the fact that the minimum energy of 
the electrons has to be about three orders of magnitudes larger than the one of the protons. In 
Fig.\ \ref{fig:GalConDiff} the influence of a higher 
electron minimum energy is shown.

\begin{figure}
\centering{
\includegraphics[height=\figheight]{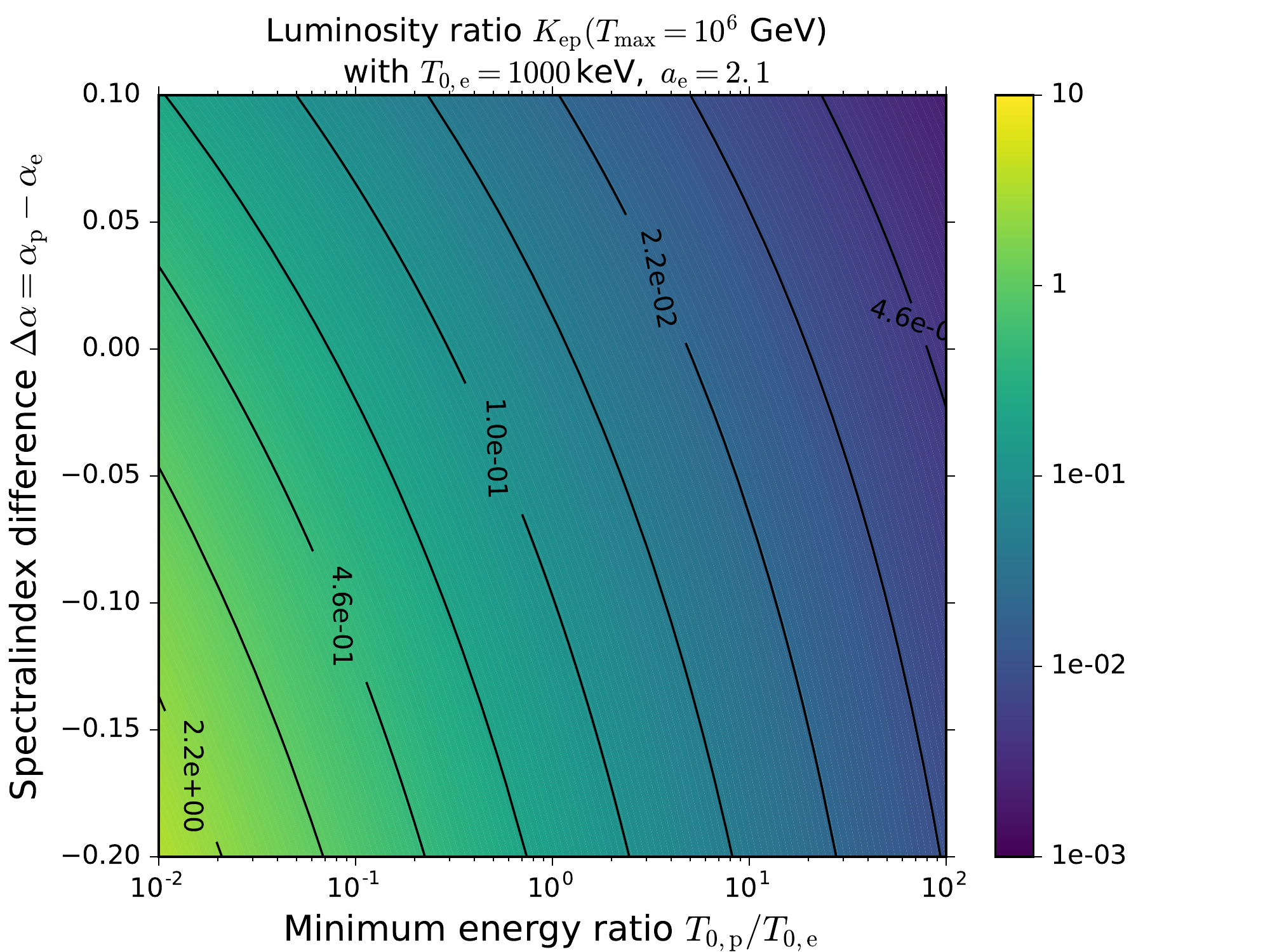}
\caption{It is clear that different minimum energies have an impact on the luminosity ratio for 
all combinations of spectral indices. With increasing $T_{0,{\rm p}}/T_{0,{\rm p}}$ the influence 
of the spectral index difference $\Delta a$ decreases.}\label{fig:GalConRatio}
}
\end{figure}

In addition to the particular case described in Fig.\ \ref{fig:GalConDiff} the difference in the 
influence of the minimum energy and spectral indices is shown in Fig.\ \ref{fig:GalConRatio}. 
Here, the minimum electron energy is fixed to an arbitrary value of $T_{0,{\rm e}}=1000\,$~keV and 
the electron spectral index is set to $a_{\rm e}=2.1$. On the first axis we used the ratio of the 
minimum energies $T_{0,{\rm p}}/T_{0,{\rm e}}$ and on the second axis the difference of the spectral 
indices $\Delta \alpha=\alpha_{\rm e} -\alpha_{\rm p}$ to display the impact of these parameter 
sets. It shows that the different minimum energies have a large impact on the value of $\kep$. 
Hence, a detailed knowledge of the acceleration process and the injection condition in particular 
is needed, in order to fix the luminosity ratio.

\subsection{Differential particle number ratio $\ktildeep$}

Different calculations for astrophysical non-thermal emission make use of a delta-functional 
approach by, for instance, coupling one frequency 
in the synchrotron spectrum of electrons to a single electron energy. Such approaches require the 
differential treatment of the ratio:

\begin{equation}
\ktildeep(T_{\rm{e}}, T_{\rm{p}})=A_{{\rm ep}}\cdot \frac{T_{\rm e}+m_{\rm 
e}}{T_{\rm p}+m_{\rm p}}\cdot\frac{{(T_{\rm 
e}^2+2T_{\rm e}m_{\rm 
e})}^{-\frac{\alpha_{\rm e}+1}{2}}}{({T_{\rm 
p}^2+2T_{\rm p}m_{\rm p})}^{-\frac{\alpha_{\rm p}+1}{2}}} \label{eq:ktildeep}\quad .
\end{equation}
For sufficiently high kinetic energies $T\gg m_{\rm p}$ and $T_{\rm e}=T_{\rm p}=T$, the 
differential ratio $\ktildeep$ is proportional to the kinetic energy to the power of the difference 
of spectral indices: $\ktildeep\propto T^{\alpha_{\rm p}-\alpha_{\rm e}}$.

\begin{figure}
\centering{
\includegraphics[height=\figheight]{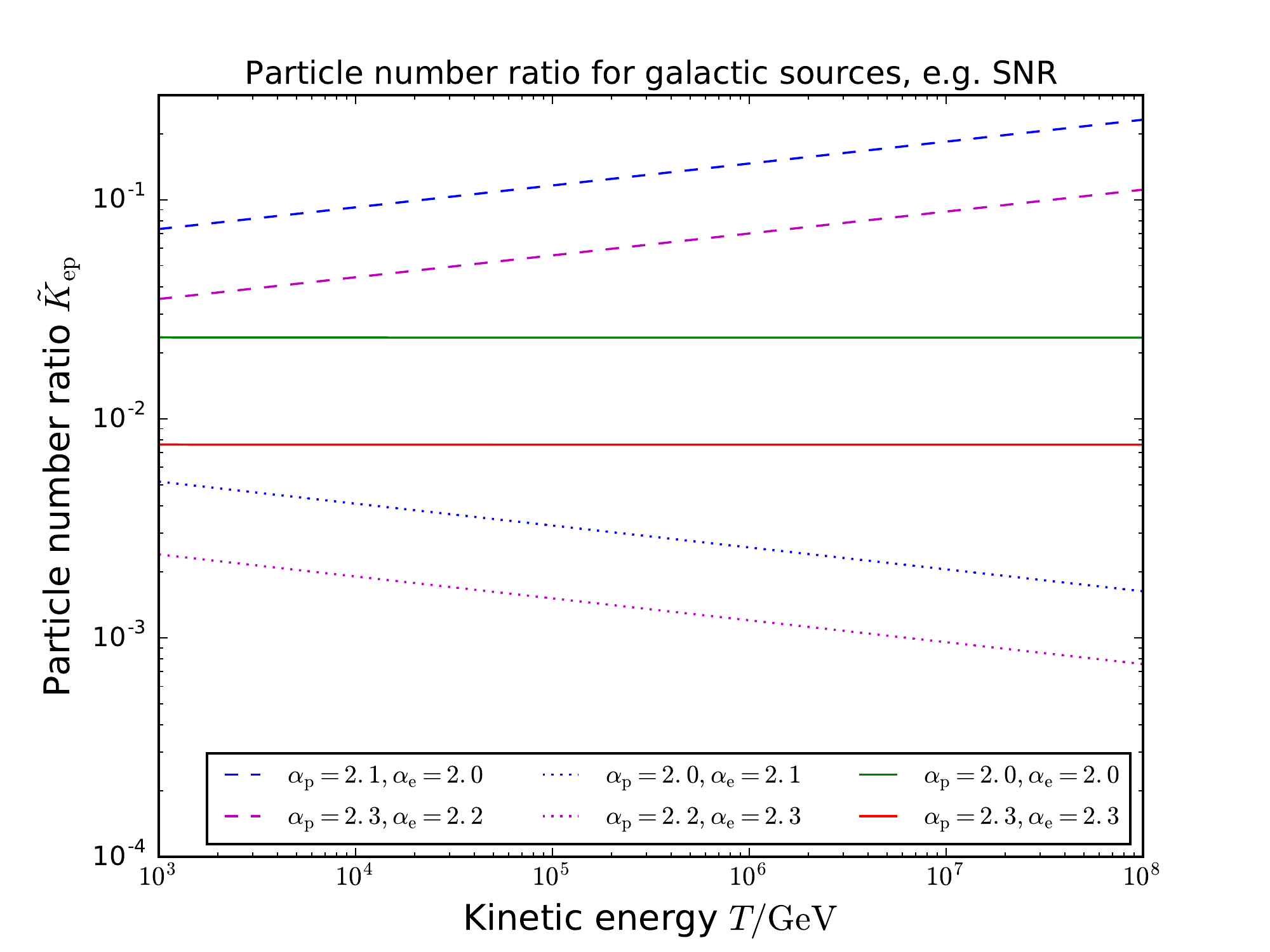}
\caption{The particle number ratio $\ktildeep$ for typical galactic 
source energies. Even for the lowest energies the asymptotic behavior 
($\ktildeep={\rm const} (\propto T^{-\Delta\alpha}))$ is already 
pronounced.}\label{ktildeep_gal}
}
\end{figure}

\begin{figure}
\centering{
\includegraphics[height=\figheight]{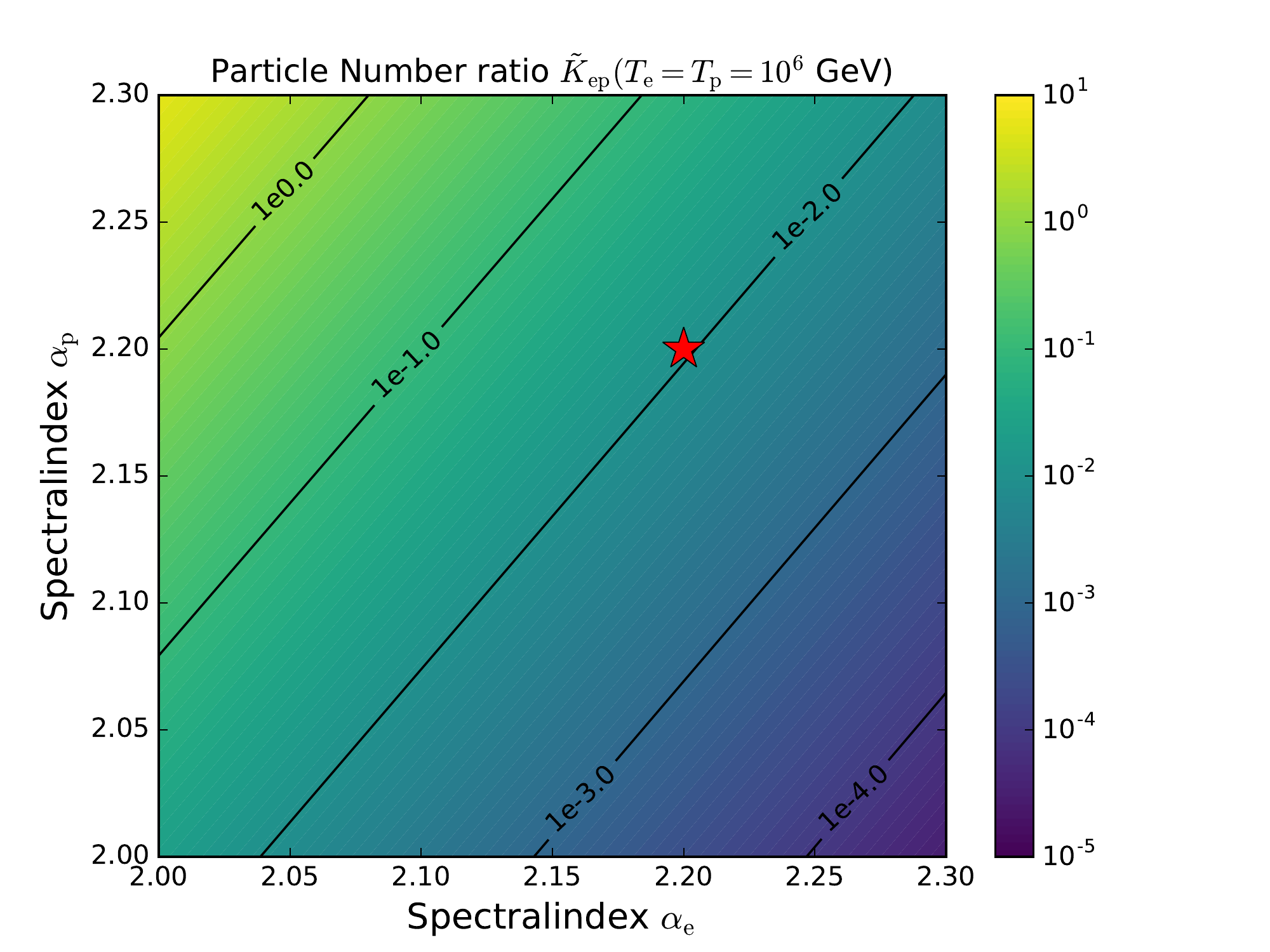}
\caption{The particle number ratio $\ktildeep$. The red star indicates the 
conventional ratio with $\alpha_{\rm e, p}=2.2$.}\label{ktildeep_contour_gal}
}
\end{figure}

\paragraph{Galactic Cosmic Rays}
Figure \ref{ktildeep_gal} shows the differential number ratio as a function of energy for galactic 
sources and different spectral indices. Again, the result is independent of energy for equal proton 
and electron spectral indices, but it varies with the exact chosen value of the index. In case of a 
differential treatment, the range of values is $0.008<\ktildeep<0.023$. The behavior becomes energy 
dependent when allowing for different proton and electron spectral indices. First we analyze the 
conservative case of a fixed index difference $\Delta\alpha=-0.1(+0.1)$, where the value of 
$\ktildeep$ varies in the range $0.035 (0.0008)<\ktildeep<0.23 (0.005)$, depending on the 
exact index and maximum energy. This behavior is shown in Fig.\ \ref{ktildeep_gal}.
The contour plot in Fig.\ \ref{ktildeep_contour_gal} shows the non-conservative case, when allowing 
for free indices of $2.0<\alpha_i<2.3$. 
Here a range of $3\cdot10^{-5}<\ktildeep<5.8$ is 
possible for a maximum energy of $T=10^{6}$~GeV. This range is much larger than for $\kep$, because 
of the strong energy dependence $\ktildeep \propto (T/{\rm MeV})^{\alpha_{\rm p}-\alpha_{\rm e}}$. 
This strong dependence leads to a simple approximation of $\kep$, which corresponds to the first 
order Taylor approximation of the formula \ref{eq:ktildeep}:
\begin{align}
 \ktildeep(T, T) \approx A_{\rm ep} \cdot 
\left(\frac{T}{{\rm MeV}}\right)^{\alpha_{\rm p}-\alpha_{\rm e}} \quad ,
\end{align}
which is very accurate and should be sufficient in most cases.

\begin{figure}
\centering{
\includegraphics[height=\figheight]{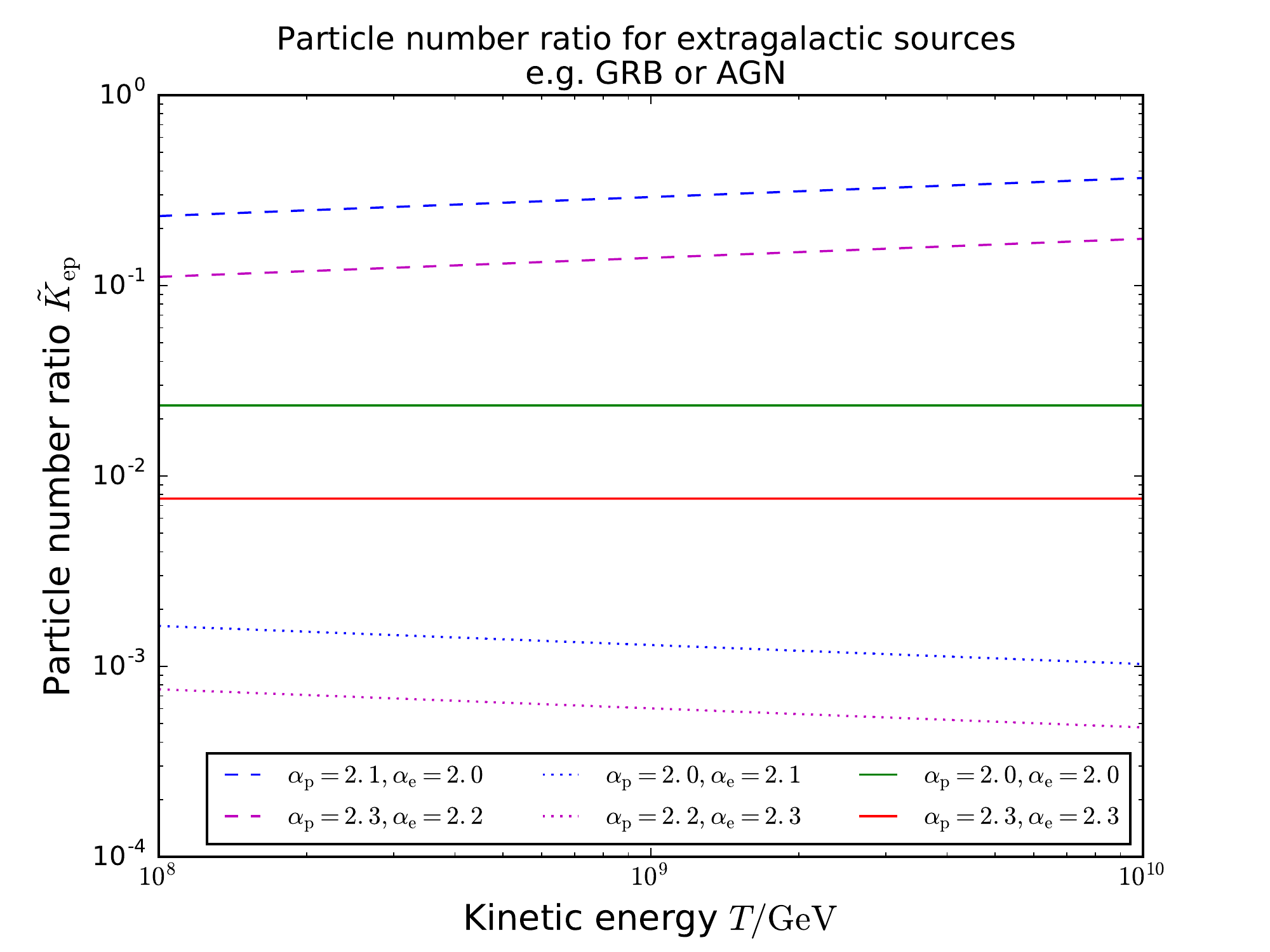}
\caption{The particle number ratio $\ktildeep$ for different maximum 
energies typical for extragalactic sources. The energy dependence is clearly visible in the case of 
differing indices}\label{ktildeep_exgal}
}
\end{figure}

\begin{figure}
\centering{
\includegraphics[height=\figheight]{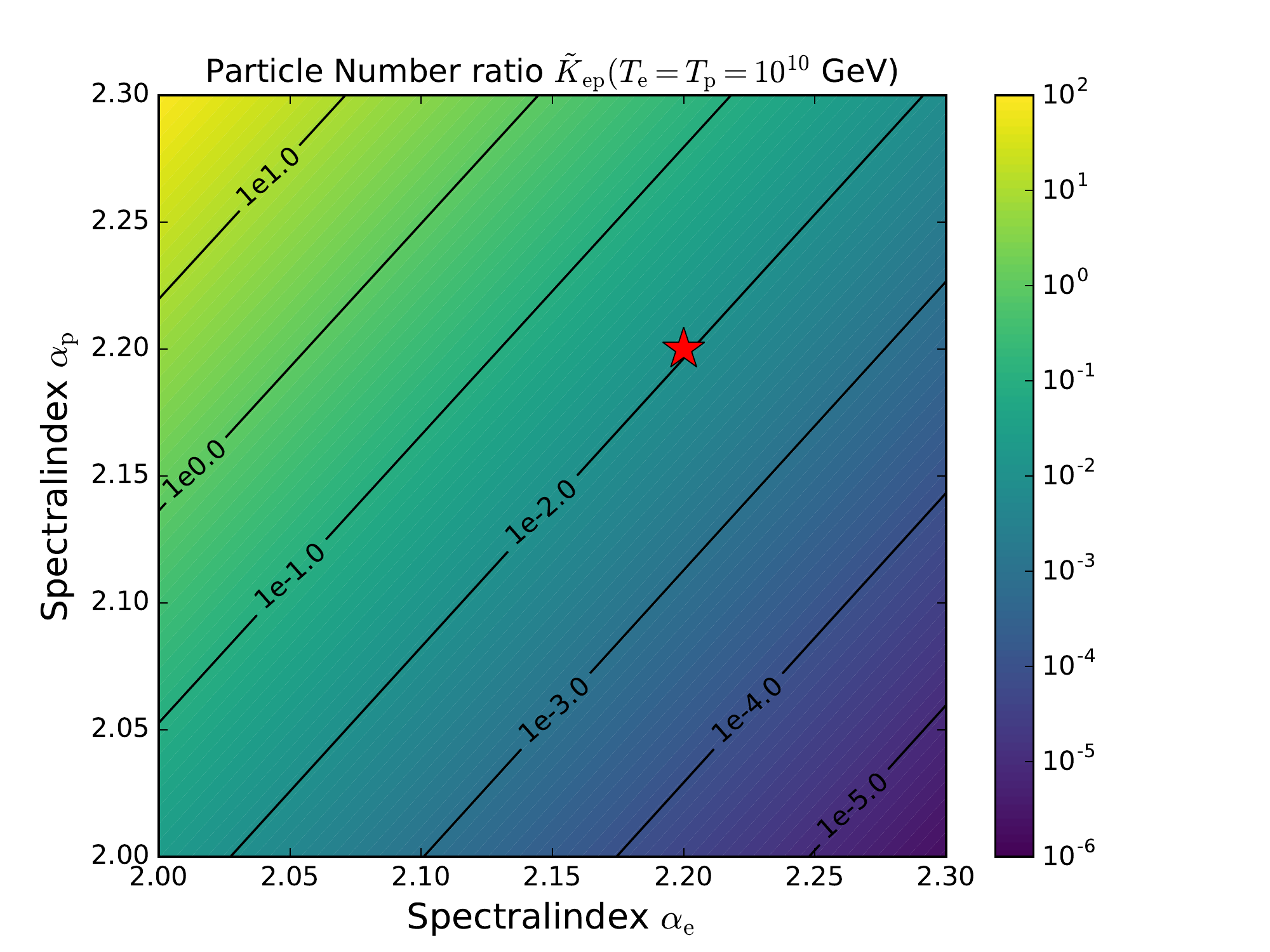}
\caption{The particle number ratio $\ktildeep$ if no restriction on 
the differences in the spectral indices are made. The red star indicates the 
conventional ratio with $\alpha_{\rm e, p}=2.2$.}\label{ktildeep_contour_exgal}
}
\end{figure}

\paragraph{Extragalactic Cosmic Rays}
The differential particle number ratio $\ktildeep$ for extragalactic sources does not differ from 
galactic sources in the case of equal indices. For slight differences in the spectral behavior of 
$\Delta\alpha = -0.1(+0.1)$ the spread of $\ktildeep$ is nearly the same as for galactic sources 
but the 
absolute value is, due to the energy dependence, higher. The graphs in Fig.\ \ref{ktildeep_exgal} 
show a variation of $\ktildeep$ in the energy range from $T=10^8-10^{10}$ GeV from $0.11 (0.0005)< 
\ktildeep < 0.37 (0.0016)$.
Finally the restriction of a fixed difference in the spectral behavior is dropped and the full 
width of theoretically allowed spectra is calculated. The results are shown in figure 
\ref{ktildeep_contour_exgal}. This figure shows that very significant differences from the well 
known value of $\ktildeep=1/100$, if one allows for slightly bigger variation in the indices. The 
values 
cover the whole range from $7.7\cdot10^{-6} <\ktildeep < 92.3 $.

\paragraph{Different minimum energies}

The influence of different minimum energies $T_{0,{\rm p}}\neq T_{0,{\rm e}}$ on the differential 
particle number ratio $\ktildeep$ is limited to the change of the normalization ratio $A_{\rm e,p}$ 
according to Eq. \ref{eq:ktildeep}. Therefore, we refer the reader to section 
\ref{ssec:Normalization} to examine the change of $\ktildeep$ with varying minimum energies since 
the general shape of the Figs. \ref{ktildeep_gal}, \ref{ktildeep_contour_gal}, \ref{ktildeep_exgal} 
and \ref{ktildeep_contour_exgal} is not changed.

\newpage
\section{Discussion of the results\label{discussion:sec}}
The central conclusion from the results above, independent on the examined source, is that there 
is a significant difference in the absolute values of the electron to proton ratio when considering 
the differential number and the integral luminosity measure. The differential number ratio is 
directly given by $A_{\rm ep}$ in the case of equal electron and proton spectral indices. The 
luminosity measure, on the other hand, is modified even for equal indices by a factor $L_{\rm 
tot}^{0}\left(m_e\right)/L_{\rm tot}^{0}\left(m_p\right)$. The change in the results is striking, 
as 
we receive values that are a factor $5$ higher for the luminosity measure as compared to the 
differential value. This is an important piece of information when trying to compare theoretical 
and 
experimental values. From astrophysical observations, the number derived often corresponds to an 
integral value. For instance, when estimating the galactic or extragalactic values for the electron 
to proton ratio, it is not possible to use differential values, as the spectral behavior of the 
cosmic rays is modified by transport. The only possibility is to use the luminosity measure, i.e.\ 
integrating over the energy range that seems relevant in the given context. The number of 
$\kep=0.01$ is derived for the Milky Way by comparing the total electron and proton luminosities at 
Earth (see e.\ g.\ the proton and electron fluxes given by the AMS collaboration 
\cite{AMS_Proton, AMS_electron}). The number of $\kep=0.1$ is derived when comparing the cosmic ray 
luminosity above the knee 
with radio emission from active galaxies or gamma-ray bursts, assuming that the radio emission 
comes 
from electrons that are co-accelerated with the hadronic cosmic rays (see e.\ g.\ 
\cite{saba2014, waxman_bahcall1997}). There is a central difficulty 
connected to this insight: the luminosity ratio bears the uncertainties of the choice of 
integration 
limits, both for the experimental number and the theoretical calculation. And as the energy spectra 
compared in observations contain two large uncertainties for the estimation of the luminosity 
ratio: 
(1) the spectra are modified by transport effects (mostly losses by radiation processes and/or 
diffusion) and this also effects the energy range; (2) the hadronic cosmic ray spectrum is a 
mixture 
of different sources, it is clearly composed of both a Galactic population and an extragalactic 
one. 
In particular for the estimate of the extragalactic population (or even populations), it is 
important to know how much these sources contribute towards low energies, as it is the lower 
integration threshold that determines the total luminosity.

Further, there are theoretical calculations that actually request the differential number ratio 
$\ktildeep$, but instead the value derived from the observational data, i.e.\ the luminosity 
ratio $\kep$, is used. In the following paragraphs, we will discuss the different source classes 
and 
how our 
findings effect the calculations that rely on the electron to proton ratio.

\subsection{Galactic cosmic rays}

In \cite{snr_matthias}, a sample of 24 remnants
with spectral information from radio up to $>$~GeV energies has been
examined statistically, with a large fraction of sources being likely
to have a dominant hadronic signature at high gamma-ray energies
\cite{snr_matthias}. Here, the hadronic spectra
tend to show spectral indices around $\alpha_p\approx 2.3$, while
the electron spectra scatter around $\alpha_e\approx 2.0$. This result is striking, as it is 
expected that the electron spectra should rather be steeper than the proton spectra due to the 
larger energy loss processes that the electrons suffer from. Standard particle acceleration theory 
suggests the same spectral behavior for protons and electrons. However, the experimental data cited 
above refer to different momenta where already very tiny deviations from the standard theory may 
imply such differences. A non-uniform behavior of the spectral indices is supported by 
observations, e.g.\ the CREAM data \cite{CreamData}. Furthermore, this sample
only represents a subclass of well-identified SNRs with a broad range of spectral indices. In 
addition the observed spectra are modified through loss processes. Despite these facts it is shown 
that a treatment of equal spectral indices might not work for individual 
sources.
Furthermore, the temporal development of supernova remnants (see e.g.\ 
\cite{cox1972,cristofari2013})
results in a change in maximum energy of the non-thermal particle spectra $T_{\max}=T_{\max}(t)$. 
According to the findings in this paper, this should also change the electron to proton ratio. 
However, this change might be negligible when compared to the uncertainties in the determination of 
the ratio from the electron and proton spectral indices.

The derivation of the luminosity ratio from astrophysical data has to be taken 
with care, since it is determined from the ratio of the diffuse flux of electrons and 
protons in 
the Galaxy, i.e.\
\begin{equation}
\kep=\frac{L_{\rm e, obs}}{L_{\rm CR, obs}}
\end{equation}
There are two central uncertainties in this calculation:
\begin{itemize}
\item The proton luminosity is determined above 1 GeV, as the spectrum is modified by ionization 
losses in the interstellar medium and by solar modulation at lower energies. In theory, 
acceleration 
is effective from above $T_{0}\approx 10$ keV , which we also use in our calculation. As 
the cosmic ray spectrum decreases with energy as $\approx E_{\rm CR}^{-2.7}$, the integral 
luminosity behaves as $E_{\min}^{-0.7}$. For protons, below one GeV, there is a break in the 
spectrum, as the power-law behavior is in momentum and not in kinetic energy, so the contribution 
that is added will increase weaker than $E_{\min}^{-0.7}$. Still, an increase 
of the proton luminosity by a factor of a few can be expected.
  \item The electron luminosity observed at Earth is highly loss-dominated: electrons lose large 
parts of their energy to synchrotron, inverse compton and bremsstrahlung losses on their way to 
Earth. This implies that part of the energy budget of cosmic electrons is lost to the thermal pool. 
It is therefore expected that the total non-thermal electron luminosity of the Galaxy is 
significantly higher than observed at Earth, which increases the electron to proton luminosity 
ratio.
\end{itemize}

\subsection{Magnetic fields and starburst galaxies}
Many authors try to derive the magnetic field strength of other galaxies. Starburst regions are of 
special interest because of their large star forming rate. These environments allow for example to 
probe the role of magnetic fields in the early universe \cite{Adebahr2013}. An example for the 
derivation of magnetic fields in nearby starburst galaxies is given by 
\cite{Adebahr2013} and \cite{Heesen2009}. They calculate the mean magnetic field strength in M 82 
and NGC 253 from observational data.
In order to derive the magnetic field strength from the observed synchrotron emission spectra of 
the electron component Beck \& Krause provide  in \cite{beck_krause2005} formulas for the 
equipartition magnetic field in a galaxy, depending on the differential number ratio:
\begin{equation}
B_{\rm eq} = \left\{\frac{4\pi(2\alpha+1)\cdot (\ktildeep+1)\cdot I_{\nu}\cdot E_{\rm 
p}^{1-2\alpha}\cdot \left(\frac{\nu}{\nu_{0}}\right)^{\alpha}}{(2\alpha-1)\cdot b(\alpha)\cdot l 
\cdot (2/3)^{(\alpha+1)/2}}\right\}^{(\alpha+3)^{-1}}\,.
\end{equation}
Here, $\alpha$ is the synchrotron spectral index, which is connected to the electron spectral index 
as $p_e = 2\alpha+1$. $I_{\nu}$ is the synchrotron intensity, $\nu_0$ is a reference frequency and 
$b(\alpha)$ is a function that depends weakly on $\alpha$. Both parameters are defined in 
\cite{beck_krause2005}, and the exact definitions are not important in this context as we want to 
discuss the dependence on $\ktildeep$, which is $B_{\rm eq}\propto 
\left(\ktildeep+1\right)^{(\alpha+3)^{-1}}$. For typical values of $\alpha\approx 0.5-1$, the 
behavior is close to $B_{\rm eq}\propto \left(\ktildeep+1\right)^{1/4}$. For values between 
$\ktildeep \approx 0.01 - 0.1$, which we find to be realistic, the difference in the calculation is 
2\% or less. Thus, any possible change in the ratio due to a more propper description of the 
underlying theory is negligible. However, if very extreme differences in the spectral indices of 
electrons to protons should be present for some reason, for instance $\Delta \alpha =\pm 0.3$, the 
differential number density varies between $10^{-5}<\ktildeep<10$ for galactic sources. In that 
case, the results change by over 50\%. There are no indications at this point that the differences 
in the spectra can be that large, but it cannot be excluded at this point either.

Lacki \& Beck \cite{lacki_beck2013} revised the above calculation for those galaxies that are 
loss-dominated, i.e.\ where the observed synchrotron radiation predominantly comes from electrons 
produced in cosmic ray interactions with the interstellar medium. This is the case for starburst 
galaxies with high cosmic ray fluxes and high densities. In the calorimetric case, the primary to 
secondary ratio $\kappa = n_{\rm sec.}/(n_{\rm sec.} + n_{\rm prim.})$ becomes $1$. This means it 
is independent on any theoretical argument of how electrons are accelerated with respect to protons.

However, Eichmann \& Tjus \cite{eichmann_beckertjus2016} used recently a more rigorous description 
of the transport and loss processes in starburst regions. Taking diffusion and advection as well as 
several different loss processes for protons and electrons into account they could show that 
some reasonable parameter configurations are dominated by primary electrons. Although most, 
including the best-fit, models show a domination by secondary electrons a dominating primary 
electron component is not ruled out. These arguments suggest a treatment of $\ktildeep$ as 
described above also in the case of starburst regions, especially for configurations with low 
target densities (up to $n_{\rm target}=1/{\rm cm}^3$).

\subsection{Extragalactic sources of ultrahigh energy cosmic rays}
\label{ssec:ExtragalacticSources}
Since we have now strong evidence that M82 is the source of the TA hot spot \cite{Abbasi2014, Fargion2015} one may  take tidal disruption events in starburst galaxies as possible sources of UHECR into account. Furthermore, Gamma-ray bursts (GRBs) have long been discussed as one of the few source classes being able to 
surpass the Hillas-limit for $10^{20}$~eV cosmic rays, necessary to explain the cosmic ray flux 
above the ankle \cite{waxman_bahcall1997}.  Recently, strong limits on the parameter space of the
average Lorentz factor $\Gamma$ of GRBs and the electron to proton luminosity fraction $\kep$ 
\cite{icecube_nature_grbs}. Here, the model of photohadronic neutrino production as first presented 
in \cite{waxman_bahcall1997} and later refined in e.g.\ \cite{guetta2004,becker2006,icecube22} was 
used in order to derive limits to the above mentioned parameter space from the fact that no 
neutrinos were observed in an analysis of 117 bursts detected at gamma-ray energies (keV-MeV 
region). The photohadronic neutrino production model in GRBs was further revised in 
\cite{winter_prl}, where it was shown that a more detailed treatment of the particle physics lead 
to a generally lower diffuse flux. For all these calculations, the electron to proton fraction was 
fixed to $\kep=1/10$, as an approximate value derived from observations. As we have shown in this 
paper, this value is well supported by theory, {\it if} the spectral indices of protons and 
electrons are only slightly different. If they are the same, the ratio will rather be smaller, 
$\kep 
\sim 0.01$, which actually would enhance the neutrino flux. One of the arguments to use $\kep 
\approx 0.1$ for the calculation of extragalactic sources is that this number is supported by 
observations. To derive this number, the luminosity of the cosmic ray spectrum above the ankle is 
compared to the total radio luminosity of the object, i.e.\ $L_{\rm CR}(E_{\rm CR}>10^{9.5}\,{\rm 
GeV})/L_{\rm radio}(\nu_{\min},\nu_{\max})$. It turns out to be $0.1$ both when considering 
gamma-ray bursts and active galaxies. This number is, however, highly uncertain. The extragalactic 
cosmic ray spectrum might continue towards lower energies, which would increase the nominator in 
the 
ratio, thus increasing the ratio itself:
\begin{equation}
  \kep(\rm extragal)=\frac{L_{\rm e}}{L_{\rm CR}^{\rm tot,extrag}}
  =\eta_{\rm obs}^{-1}\cdot \frac{L_{\rm e}}{L_{\rm CR}}
\end{equation}
with $L_{\rm CR}^{\rm tot,extrag}=\eta_{\rm obs}\cdot L_{\rm CR}(L_{\rm CR}^{\rm 
tot,extrag}/L_{\rm e})$ and $\eta_{\rm obs}>1$.
On the other hand, only a fraction of the total energy in electrons might be transferred to 
synchrotron radiation, and in addition, the radio component only represents a part of the total 
energy, i.e.\ $L_{e}=\chi(B,\nu_{\min},\nu_{\max})\cdot L_{\rm radio}>L_{\rm radio}$. Thus, the 
luminosity ratio becomes
\begin{equation}
\kep(\rm extrag)=\frac{\chi(B,\nu_{\min},\nu_{\max})}{\eta_{\rm obs}}\cdot \frac{L_{\rm CR}}{L_{\rm 
e}}\,.
\end{equation}

The same arguments that are used for GRBs are also used when determining the neutrino flux for 
active galactic nuclei (AGN), see e.g.\ \cite{becker2008} for a review. In \cite{becker2014}, 
$\chi$ 
was determined to be $\chi\approx 100$ for magnetic field values below about 1 Gauss 
\cite{becker2014}. $\eta_{\rm obs}$, on the other hand, can be equally large, as the cosmic 
ray spectrum behaves as $E^{-2.7}$ above the ankle. Thus, the cosmic ray luminosity scales with 
$E_{\min}^{-0.7}$. Reducing the minimum energy by one, two or three orders of magnitude would 
result 
in $\eta_{\rm obs}=5,\,25,\,125$, respectively and thus bringing us back to the original number.

\newpage
\section{Summary and Outlook\label{sum:sec}}
In this paper, we reinvestigate the ratio of electrons to protons from a theoretical perspective. 
In 
particular, we drop the assumption of equal spectral indices for electrons and protons and 
distinguish between the ratio of differential energy spectra, $\ktildeep$ and the luminosity ratio 
$\kep$. These two definitions are often used as equivalent in the literature. As we can show in 
this 
paper, they differ significantly from each other and it is therefore necessary to precisely define 
how the ratio is used in calculations. In this paper, we present an analytic solution 
for the luminosity ratio, which can be directly compared to the measured electron and proton 
luminosity ratio at Earth. Our results lead to the following main conclusions:
         \begin{enumerate}
          \item
The luminosity ratio is typically higher than the differential ratio, i.e.\ $\kep>\ktildeep$. This 
is 
the case even for equal electron and proton indices, as $\ktildeep$ is directly given by the ratio 
of spectral normalizations of the two species, while the luminosity ratio $\kep$ is modified by the 
ratio of the integral luminosities of the two particle fluxes. Typical values lie around $\kep \sim 
0.008-0.1$, and can thus be significantly larger than the value that is  typically assumed for 
sources of galactic cosmic rays, i.e.\ $\kep\sim 0.01$. The differential values vary with energy if 
the spectral behavior of protons and electrons differ from each other. In the case of same spectral 
indices in the range $2.0<\alpha<2.3$, the differential ratio is $0.008<\ktildeep<0.23$, while it 
can vary significantly more for different spectral indices. For different minimum energies of 
electrons and protons $\kep$ and $\ktildeep$ change significantly. However, an equal but varying 
minimum energy up to $T_0\lessapprox 0.1\,$~MeV has a negligible influence.
            \item Those ratios of electrons to protons derived from observations typically 
represent 
luminosity measures. Even those need to be taken with care, as they are given for certain energy 
ranges, where cosmic ray phenomena are observable. As an example, the flux of extragalactic hadronic 
cosmic rays is detected above $3\cdot 10^{18}$~eV, where it is quite clear that the flux must come 
from extragalactic sources. This energy value is usually used as a lower integration threshold when 
deriving the required electron to proton ratio of extragalactic sources. However, it is likely that 
the extragalactic component continues toward lower energies, even if it might be subdominant 
compared to the Galactic component at these energies, it will still contribute significantly to the 
complete electron to proton ratio. Thus, theoretical and experimental electron to proton ratios can 
only be compared when taking these uncertainties into account.
\item In particular, calculations of neutrino fluxes for neutrino emission from gamma-ray bursts and 
active galaxies rely on the estimate of the luminosity ratio. The typical value used here is $\kep = 
0.1$ as derived from observations. This number is highly uncertain in both directions as discussed 
above. From theoretical arguments as presented in this paper, we would rather expect it to be a 
factor of a few smaller if electron and proton indices are the same, i.e.\ around $\kep \approx 
0.01$, which would increase the expected neutrino flux with respect to the presented calculations 
(e.g.\ \cite{guetta2004,becker2006,Huemmer2012,becker2014}).
\item The calculation of magnetic fields from synchrotron depends on the differential ratio $\ktildeep$ \cite{beck_krause2005}. As the dependence is relatively weak, a change in the number from the typical value of $\ktildeep=0.01$ only becomes significant when allowing for strong differences between the spectral behavior of electrons and protons. A difference in the spectral index of $\Delta \alpha=0.3$ would result in a $\approx 50\%$ change in the equation to determine the B-field. In sources where the secondary electron population from cosmic ray interactions becomes dominant \cite{lacki_beck2013}, the effect should be negligible.
            \end{enumerate}

In the future, these results can be applied to concrete sources. For those cases, the absolute 
proton and electron luminosities can be used as an additional criterion to further constrain a 
possible range of values, as each individual source carries a maximum energy budget that actually 
can be transferred to cosmic rays. We consider this paper as a first, general study of the phenomena 
that can be applied more concretely in the future to individual galactic and extragalactic sources, 
but also when considering the investigation of source populations.

\newpage
\section*{Acknowledgments}
The authors would like to thank Matthias Mandelartz, Micha{\l} Ostrowski, Isaac Saba, Florian 
Schuppan and Walter Winter for valuable discussions. Furthermore, the authors would like to 
thank the anonymous reviewer for the valuable comments. This work is supported by the 
DFG-research 
unit {\it FOR1048, ``Instabilities, turbulence and transport in cosmic 
magnetic fields''} and the RAPP Center (Ruhr Astroparticle and Plasma Physics Center), funded by 
the 
MERCUR project St-2014-040. LM was partially funded by the Deutsche Studentenf\"orderung of 
the Konrad Adenauer Stiftung. Further support comes from the research department of plasmas with 
complex interactions (Bochum).

\newpage
\appendix

\section*{References}
\bibliography{lib}

\end{document}